**Automatic Detection of Cortical Arousals in Sleep and their Contribution to Daytime Sleepiness**

Andreas Brink-Kjaer[a,b,c], Alexander Neergaard Olesen[a,b,c], Paul E. Peppard[d], Katie L. Stone[e,f], Poul Jennum[c*], Emmanuel Mignot[a*], Helge B.D. Sorensen[b*]

*Shared last author

[a] Center for Sleep Science and Medicine, Stanford University, California, USA

[b] Department of Health Technology, Technical University of Denmark, Kongens Lyngby, Denmark

[c] Danish Center for Sleep Medicine, Glostrup University Hospital, Glostrup, Denmark

[d] Department of Population Health Sciences, University of Wisconsin-Madison, Madison, Wisconsin, USA

[e] Research Institute, California Pacific Medical Center, San Francisco, CA

[f] Department of Epidemiology and Biostatistics, University of California, San Francisco

Corresponding author:

Name: Andreas Brink-Kjaer

E-mail address: andbri@dtu.dk

**Abstract**

Cortical arousals are transient events of disturbed sleep that occur spontaneously or in response to stimuli such as apneic events. The gold standard for arousal detection in human polysomnographic recordings (PSGs) is manual annotation by expert human scorers, a method with significant interscorer variability. In this study, we developed an automated method, the Multimodal Arousal Detector (MAD), to detect arousals using deep learning methods. The MAD was trained on 2,889 PSGs to detect both cortical arousals and wakefulness in 1 second intervals. Furthermore, the relationship between MAD-predicted labels on PSGs and next day mean sleep latency (MSL) on a multiple sleep latency test (MSLT), a reflection of daytime sleepiness, was analyzed in 1447 MSLT instances in 873 subjects. In a dataset of 1,026 PSGs, the MAD achieved a F1 score of 0.76 for arousal detection, while wakefulness was predicted with an accuracy of 0.95. In 60 PSGs scored by multiple human expert technicians, the MAD significantly outperformed the average human scorer for arousal detection with a difference in F1 score of 0.09. After controlling for other known covariates, a doubling of the arousal index was associated with an average decrease in MSL of 40 seconds ($\beta$ = -0.67, $p$ = 0.0075). The MAD outperformed the average human expert and the MAD-predicted arousals were shown to be significant predictors of MSL, which demonstrate clinical validity the MAD.

## 1 Introduction

Sleep dysregulation and sleep disorders are associated with cardiovascular, metabolic, and psychiatric disorders [1, 2]. Sleep dysfunction is usually evaluated in sleep clinics, through analysis of nocturnal polysomnography (PSG) and, less frequently using multiple sleep latency tests (MSLTs), an objective measure of daytime sleepiness where latencies to sleep are measured during 4-5 daytime naps. A PSG recording involves measuring electroencephalography (EEG), electrooculography (EOG), electromyography (EMG), electrocardiography (ECG), airflow, respiratory effort and blood oxygen saturation during a night's sleep. PSGs are conducted at night, and result in the scoring of Sleep Disordered Breathing (SDB) events (reported number of apneas or hypopneas per hour of sleep as the Apnea Hypopnea Index, or AHI), periodic leg movements during sleep (number of periodic leg movement (PLM) event per hour of sleep with and without associated arousals, PLMI and PLMAI respectively), and sleep stages [wakefulness (W), non REM sleep (stage one N1, two N2, or three N3) or REM sleep (R)], reported as percent of total sleep time. AHI can be reported either as “recommended” by the American Academy of Sleep Medicine (AASM), which includes only hypopneas associated with 4% oxygen desaturation; or the AASM “alternate” AHI which counts apneas associated with 3% oxygen desaturation and/or arousal. A typical sleep study also reports sleep latency, latency for sleep onset to the first epoch of REM sleep, and sleep efficiency (SE), the percent of time asleep when in bed. A slight variation of sleep efficiency is Wake After Sleep Onset (WASO), which, unlike SE, only considers wake after sleep onset has occurred. In the context of sleep stage scoring, sleep stages are attributed to successive 30 second epochs using a majority rule, an arbitrary decision historically justified by the use of paper printing in sleep studies [3].

In addition to traditional sleep scoring every 30 seconds, sleep is often disturbed by transient arousal microevents between 3 and 15 seconds, or smaller wake segments that disturb the EEG, but that are not long enough to be scored as a full epoch of wakefulness (>15 sec of wake during a 30 sec epoch). These events, called microarousals, are also often associated with brief increases in muscle tone in the EMG, another important feature allowing proper scoring of arousals notably during REM sleep. Although microarousals can occur naturally as part of normal sleep-wake physiology, these are often the result of external stimuli (e.g. disturbing sound) or internal sleep disorder events such as SDB (i.e. sleep apnea) or PLM events. In traditional sleep study scoring, arousals are not always systematically scored, although in most instances arousals, whether of short duration or resulting in a full-blown epoch of wakefulness, are generally scored as part of the alternate AHI and the PLMAI. Scoring of cortical arousals in this context is carried out according to AASM guidelines [4].

As mentioned above, wake events disturbing sleep longer than 15 seconds are reported as part of SE or WASO, whereas microarousals are only systematically scored by technicians when following SDB or PLM events. An excessive number of arousals, whether in the form of brief epochs of wakefulness (integrated in SE and WASO measures), or as microarousals (reported as part of the alternate AHI or PLMI), is associated with sleep fragmentation and poor sleep, which in turn is linked to daytime sleepiness [5]. Subjects with excessive daytime sleepiness have a sevenfold greater rate of automobile accidents [6] and decreased quality of life.

As of today, the gold standard for detecting arousals is through visual inspection of PSG recordings [5]. This approach is both time-consuming, expensive, and has a low intra- and interscorer reliability due to subjective interpretation of used scoring guidelines [7]. Furthermore, as mentioned above, many shorter-duration arousals are often only scored in the context of SDB or PLMs, which is strongly limiting as spontaneous arousals are also likely affecting physiology.

The limitations of current arousal detection approaches have motivated the development of algorithms to automatically detect arousals, notably microarousals. De Carli (1999) published a method for detecting arousals with a recall of 0.88 by processing features of EEG and EMG through a linear discriminant [8]. Sugi, Kawana, and Nakamura (2009) developed a method using thresholds for detection of arousals associated with SDB with a recall of 0.86 [9]. Cho et al. (2006) used a machine learning approach with a support-vector machine (SVM) classifier that achieved a recall of 0.75 [10]. Popovic et al. (2014) proposed a system for detection of arousals based on a single channel EEG (Fp1-Fp2) with a recall of 0.72 and a precision of 0.67 [11]. Shahrbabaki et al. (2015) proposed a method that also included ECG recordings, leg EMG, and respiratory traces to model arousals with *k*-nearest neighbors (KNN) classifier with a recall of 0.79 [12]. Sorensen et al. (2012) used feed-forward neural network to classify arousals with a recall of 0.89 and a precision of 0.86 [13]. Fernández-Verala et al. (2017) proposed an approach for detecting arousals using a simple model based on thresholds and achieved a recall of 0.75 and precision of 0.86 [14]. A method proposed by Shmiel et al. (2009) [15] makes use of sequential pattern discovery for detecting arousals with a recall of 0.75 and precision of 0.77.

These studies all provide working arousal detection systems, however, they have many limitations. First, these detectors have only been validated in small datasets that have been manually scored by only a few human scorers, which makes comparison and generalization difficult to assess. Second, although Coppieterst Wallant et al. (2016) showed that performance of these automatic detectors was in line with inter-scorer variability, this was demonstrated using six independent scorers based on only four PSGs [16]. Third, the gold standard for scoring arousals through visual inspection of PSG is based on rules of duration that distinguishes microarousals (3-15 seconds) and wake (>15 seconds), a distinction which is arguably arbitrary.

In this study, we aimed at developing a fully automatic system, the Multimodal Arousal Detector (MAD), for the detection of all microarousals and wake events using recent advances in machine learning such convolutional and recurrent neural networks. The proposed approach challenges the gold standard by combining automatic scoring of arousals and wake with a 1-second resolution as a single measure. Furthermore, our study also differs from previously reported state-of-the-art methods given the much larger sample size, as 5,362 PSGs gathered at multiple locations are used. Furthermore, to validate our automated measures and show clinical validity, arousal detection metrics computed by the detector during nocturnal PSGs were studied as predictors of daytime sleepiness, as assessed by the MSLT.

## 2 Method

### 2.1 Data Description

Diversity in datasets is a prerequisite for developing and validating deep learning detectors [17]. Sufficient data diversity was ensured by using data from thousands of subjects from four cohorts, scored by different sleep technicians based at various sleep centers. These include the MrOS Sleep Study [18, 19, 20, 21], the Cleveland Family Study (CFS) [18, 22, 23], the Wisconsin Sleep Cohort (WSC) [24, 25], and the Stanford Sleep Cohort (SSC) [26]. All PSGs included at least a central EEG derivation, left and right EOG, chin EMG, and lead II ECG. The cohorts contain in-lab recordings (WSC, SSC), as well as Home Sleep Testing (HST) (MrOS, CFS). Signals were sampled at frequencies between 100 and 512 Hz. We note that not all cohorts had consistent and systematic scoring of all microarousals. To test interscorer reliability in comparison to detector performance, PSGs from 30 SSC and 30 WSC subjects were annotated five times each using a pool of nine sleep technicians. Annotations included microarousal scoring as defined in the AASM manual [4]. The subset of 30 SSC PSGs comprised patients with various sleep disorders such as SDB ($n$=24), insomnia ($n$=4), delayed sleep phase syndrome ($n$=1), and others ($n$=4).

#### 2.1.1 MrOS Sleep Study

The MrOS Sleep Study is a multi-center community-based cohort designed to study the relationships between sleep disorders and vascular disease, falls, fractures, and mortality in older men. A total of 2,909 men age 67 years or older underwent a full unattended PSG at six clinical sites in the United States [18, 19, 20, 21]. PSG recordings were acquired with Compumedics Safiro Sleep Monitoring System that used a high pass pre-filter with a cut-off frequency of 0.16 Hz. A total of 2,888 PSGs were included from this study. All arousals (not just those associated with SDB or PLMs) were scored according to an older ASDA (American Sleep Disorders Association's) definition [27, 21], precursor of the current AASM definition. Sleep stages were scored based on Rechtschaffen and Kales (RK) rules, however rules were slightly modified (e.g. deep sleep was scored as N3 sleep) [28]. As deep sleep (N4) was scored as N3, the difference in sleep stage scoring guidelines is considered negligible.

#### 2.1.2 Cleveland Family Study

The CFS is a large family study of sleep apnea conducted in 2,284 subjects of age between 6 and 88 from 361 families [18, 22, 23]. PSGs in this study were recorded using a Compumedics E-Series System and a band-pass pre-filter with cut-off frequencies of 0.16 Hz and 105 Hz. This study included 726 PSGs from the CFS. Certified sleep technicians scored all PSGs with rules similar to those used in the MrOS Sleep Study. All arousals were scored in the CFS sample.

### 2.1.3 Wisconsin Sleep Cohort

Participants of the WSC study were randomly sampled from Wisconsin state agencies [24, 25], the age of the participants that was included ranged from 37 to 85. PSGs in this study were measured with a Grass Comet Lab based system using a pre-filter with cut-off frequencies 0.3 Hz and 35 Hz for EEG and EOG, while cut-off frequencies of 10 Hz and 70 Hz was used for EMG [25]. Among the large WSC sample, a subset of 271 PSGs had all arousals scored, although only the onset of arousals was annotated. In addition, 1,447 PSGs with an associated MSLT, but without independent scoring of arousals were included. In this study, these data were used to examine the clinical significance of the proposed system by comparing statistics of model predictions to daytime sleepiness.

### 2.1.4 Stanford Sleep Cohort

PSGs in the SSC were recorded in patients with a wide range of sleep disorders at the Stanford Sleep Clinic [26]. PSGs in the SSC were recorded using a band-pass pre-filter with cut-off frequencies of 0.1 Hz and 0.45 times the sampling frequency. Thirty subjects of age between 20 and 90 had all arousals scored.

### 2.1.5 Data Usage and Summary

In developing the MAD, one dataset was used as a training set while a separate test dataset was set aside to provide unbiased estimates of model's performance. PSGs from MrOS and CFS were the only datasets with accurate and consistent scoring of all arousals. We therefore used 80 % of randomly selected subjects from both cohorts as the training set, while keeping the remaining 20 % for testing. Also included in the validation set were 271 PSGs from the WSC which had arousal scoring consistent with MrOS and CFS.

The remaining PSGs from WSC ($n$=1447) did not have the necessary arousal scoring for inclusion in the validation set but did have associated MSLT scores. These data are used to test whether MAD-predicted arousals are related to objective sleepiness measured clinically by the MSLT.

Finally, a separate test set of 30 PSGs from each of WSC and SSC was used to compare model performance to the performance of human scorers. The demographics as well as data usage are described in Table 1. Subjects are aged 6 to 90 and are generally overweight with an average BMI of 30.4. Research performed in this manuscript has been reviewed and approved by Stanford Institutional Review Boards.

| Name | Age (μ ± σ) | BMI (μ ± σ) | Sex (% Male) | PSGs (subjects) | | |
|---|---|---|---|---|---|---|
| | | | | Arousal Scoring | | Sleepiness Association Statistics |
| | | | | Training | Testing | |
| **MrOS** | 76.4 ± 5.5 | 27.2 ± 3.8 | 100 | 2308 | 580 | - |
| **CFS** | 41.4 ± 19.3 | 32.4 ± 9.5 | 44.8 | 581 | 145 | - |
| **WSC** | 60.0 ± 8.5 | 31.7 ± 7.2 | 53.8 | - | 271 (269) | 1447 (873) |
| **SSC** | 53.5 ± 15.8 | 29.1 ± 8.8 | 66.7 | - | 30 | - |
| **Total** | 65.4 ± 15 | 30.4 ± 7.1 | 74.3 | 2889 | 1026 (1024) | 1447 (873) |

**Table 1**: Summary of data demographics and data split. MrOS: MrOS Sleep Study, CFS: Cleveland Family Study, WSC: Wisconsin Sleep Cohort, SSC: Stanford Sleep Cohort.

## 2.2 Preprocessing of Biomedical Signals

Physiological signals were acquired using varying signal montages and can be contaminated with artifacts that decrease the signal quality. To enable consistent quality, preprocessing steps described below and summarized in Fig. 1 were performed.

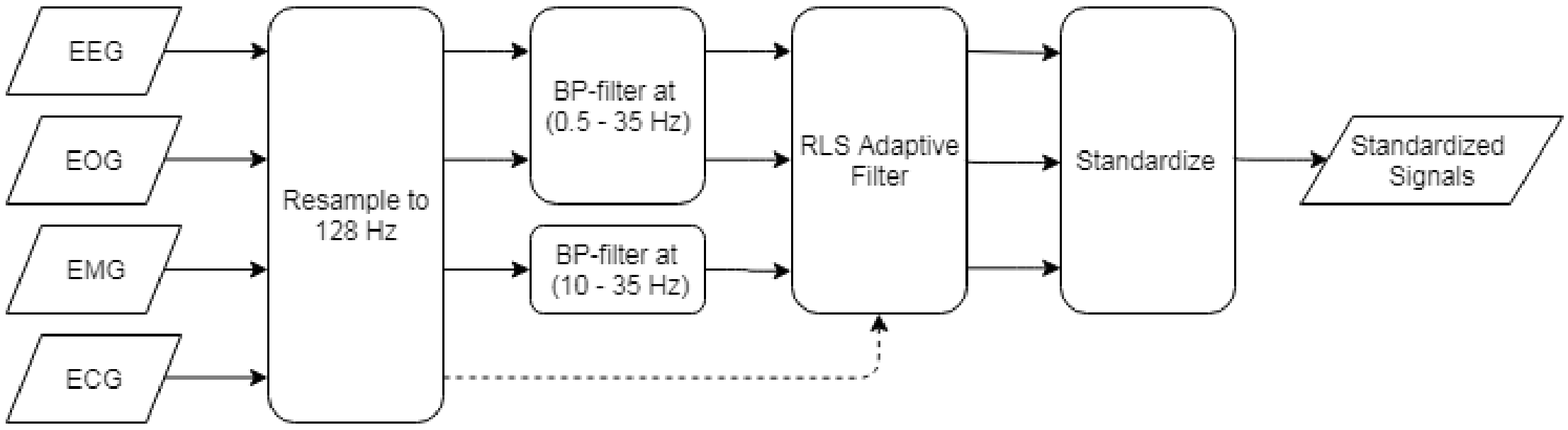

**Figure 1**: Schematic illustration of the proposed preprocessing method. Input signals are resampled, band pass filtered, the RLS adaptive filter is applied, and finally signals are standardized.

This study used convolutional neural networks that employs a series of filters to compute descriptive signal features i.e. a low dimensional representation of the signals. Filter kernels designed by a convolutional neural network are static in size, hence networks expect an input with a consistent sampling frequency. For this reason, we resampled all signals to 128 Hz. Aliasing effects were avoided by applying an anti-aliasing low pass least-squares Finite Impulse Response (FIR) filter. The FIR filter was designed to minimize the weighted integrated squared error between the filters magnitude response and an ideal piecewise function over a set of desired frequency bands that allow resampling without aliasing. Finally, a Kaiser window with a shape factor $\beta = 5$ was used to normalize the filter gain.

Infinite Impulse Response (IIR) band pass filters were used to remove frequency content that represent power line interference at 60 Hz and low frequency artifacts, while preserving physiological meaningful data. The IIR filters designed for EEG and EOG had passband ranges from 0.5 to 35 Hz with an allowed passband ripple of 1 dB, and stopband frequencies of 0.1 and 50 Hz with a stopband attenuation of 20 dB using the Butterworth method. The IIR filter used for EMG was similar but had a first stopband at 5 Hz and a first passband at 10 Hz. Filters were implemented with zero-phase filtering to avoid the issue of non-linear phase response and frequency-dependent group delay.

Recursive Least Square (RLS) adaptive filters as implemented by Moore et al. (2014) [29] and He, Wilson and Russel (2004) [30] were used to respectively remove ECG and ocular movement artifacts in the EEG. Both studies reported that using a filter order of $p = 4$ and a forgetting factor $\lambda = 0.995$ resulted in satisfactory results. RLS adaptive filters with the same settings were similarly used in this study.

Prior to feeding signals to the deep neural network, signal distributions were standardized by subtracting the mean and dividing by the standard deviation. Alternatives to this approach such as scaling by data percentiles, hard normalization, and log transformation exist, however this commonly used simple standardization [17, 31] performed well for our application.

### 2.3 Classification with Convolutional and LSTM Neural Networks

Uses of convolutional and recurrent neural networks have been shown to achieve state-of-the-art performance in various fields [17, 32], including sleep analysis [3, 33, 34]. A convolutional neural network (CNN) works by taking a static input such as a signal or image, and the CNN processes it with a network of filters. In CNNs, each layer transforms an input into a feature map, thereby automating non-linear feature extraction. In 1997, Hochreiter and Schmidhuber (1997) introduced long short-term memory (LSTM) networks, a type of recurrent neural network capable of modelling long-term dependencies without the problem of exploding or vanishing gradients [35]. The LSTM network contains a cell state, which saves information about long and short-term changes of CNN input features. The information saved is controlled by a set of gates, which depending on the input features, decides what information is saved and what is discarded. This memory allows features to incorporate temporal context, such those indicating prior sleep stage or frequency shift, which is ideally suited to detect arousals.

### 2.4 Network Architecture

The architecture of the proposed network is based on similar studies of sleep staging [3, 33, 34, 36]. The general idea is to use CNNs to automatically design a set of features describing the preprocessed EEG, EOG, and EMG signals in 1 second bins. These features are then fed to a bi-directional LSTM network followed by fully connected neural networks to predict labels (*arousal*, *non-arousal*) and (*wake*, *sleep*) as probabilities for each successive second of data. The fully connected layers enable correct classification of arousal and wake by making a highly complex non-linear mapping of features computed at the level of the LSTM layers. Labels are associated with a 1 second signal bin, consisting of 4 signals with 128 samples. The proposed CNN is based heavily on the network structure used in the work of He et al. (2015). This network structure performs very well in the field of image recognition [37] and is easily restructured for 1-D inputs. The network is comprised of residual building blocks that use convolutional layers, batch normalization [38], ReLU activation functions [39], and residual learning. He et al. (2015) proposed a deep CNN with residual learning that improved performance without adding further network complexity. Residual learning is implemented as shortcut connections as identity mappings between every second convolutional layer. Identity mappings are straight forward with identical dimensions, for increasing dimensions zero-padding is used, and decreasing dimensions are handled with max pooling. A simple illustration of the network architecture and how signals are processed is shown in Fig. 2. The full network is displayed in Fig. 3. In the network, all convolutional and fully connected layers are followed by batch normalization and ReLU activation, while output probabilities are computed using the softmax activation.

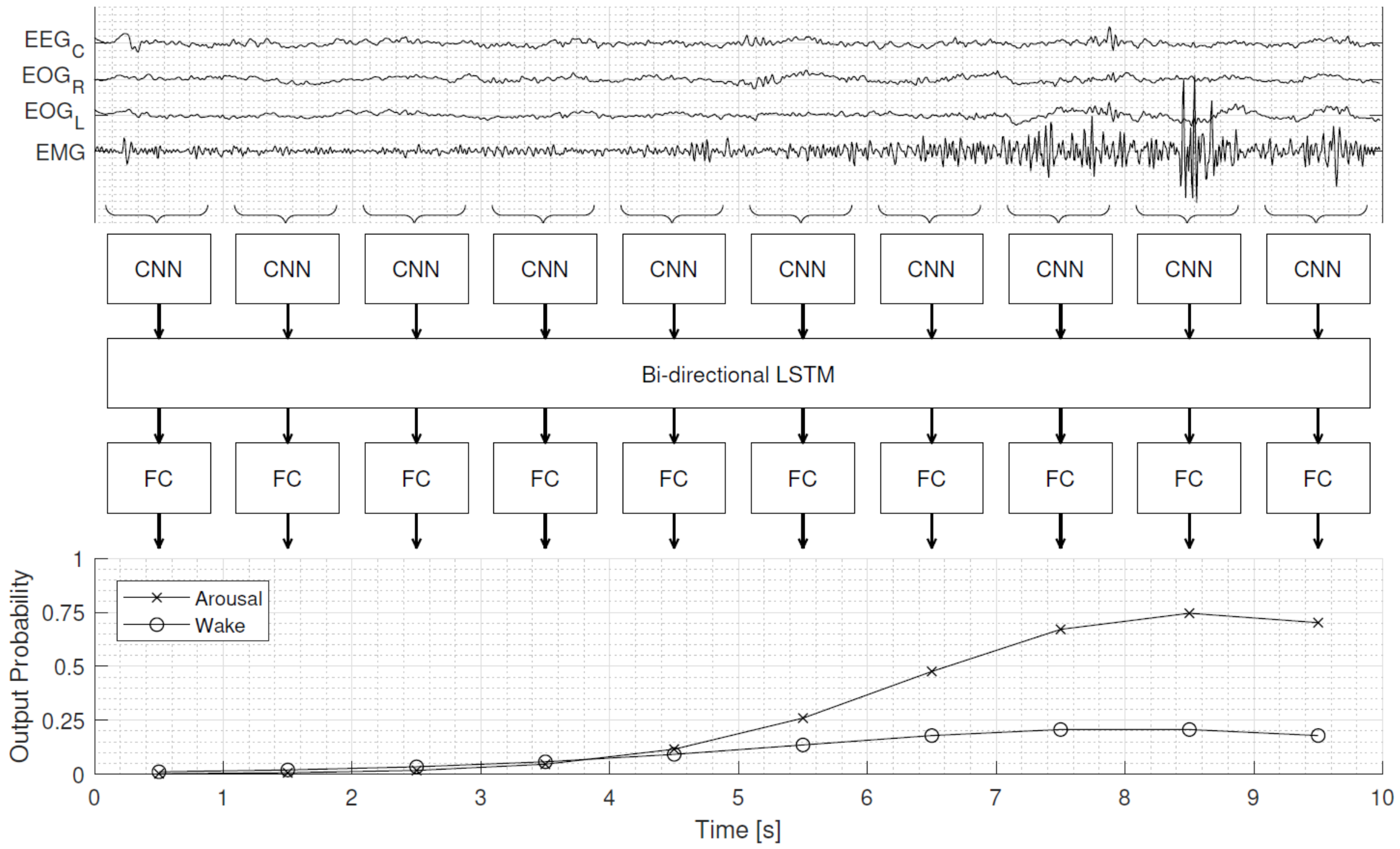

**Figure 2**: Illustration of the process to which the preprocessed signals are reshaped and fed through the proposed MAD network. The top plot shows a 10 second window of signals with an arousal annotated at 5 seconds. Arousal and wake probabilities are computed from the reshaped input being fed through the network structure consisting of CNN (convolutional neural network), LSTM (Long Short-Term Memory), and FC (fully connected) layers. The three network types (CNN, LSTM, FC) each have a main function, (1) extraction of features describing the 1 second segment of the signals, (2) computing features with temporal information (e.g. frequency change), (3) mapping temporal features to predict wake and arousal labels.

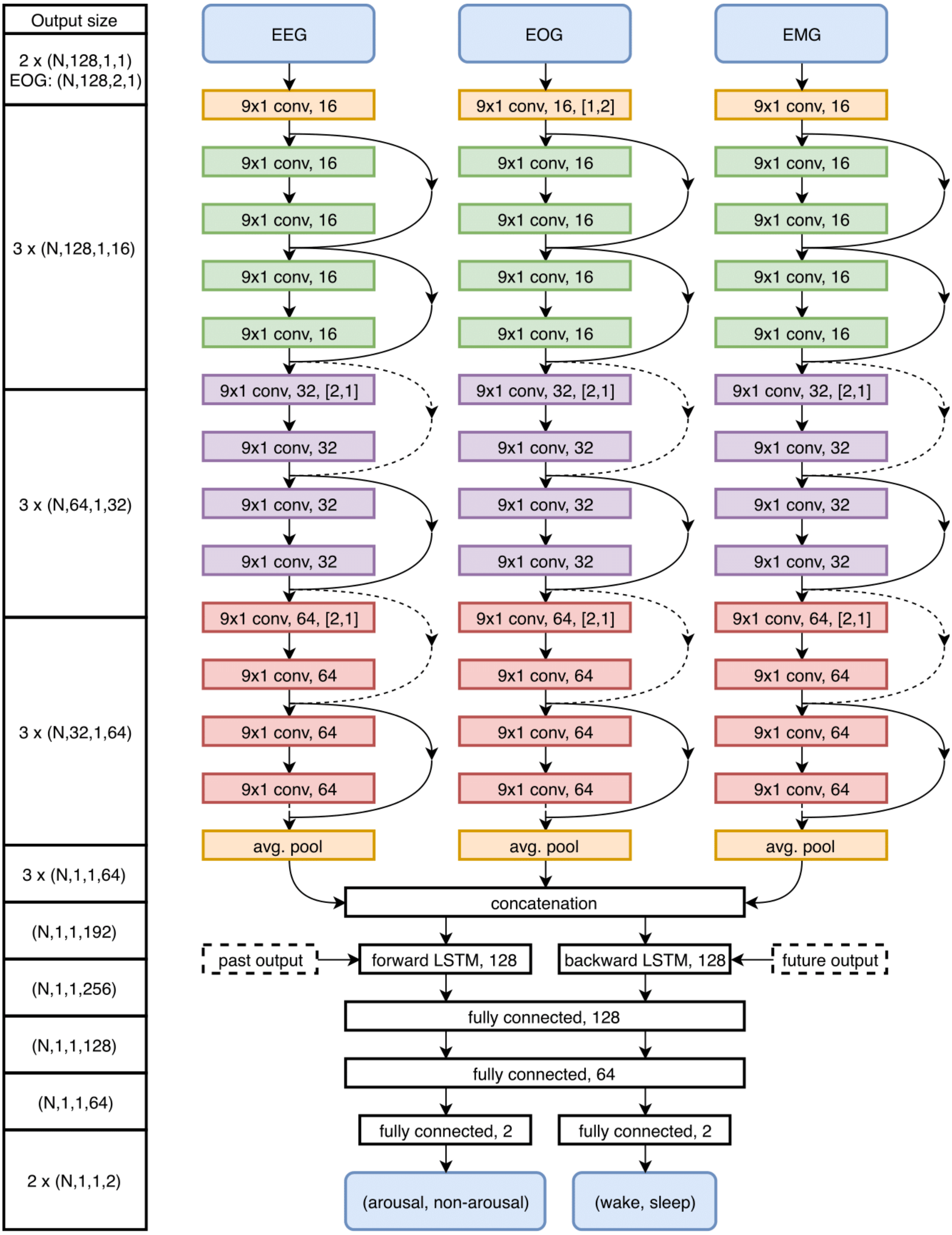

**Figure 3**: Deep neural network architecture visualization for MAD. The left column displays the output size throughout the network with dimensions defined as (N: mini-batch size, height, width, channels). Convolutional layers are described by filter kernel size, feature map size, and stride. LSTM layers are described by the number of cells. Fully connected layers are described by the number of hidden units. Shortcut connections are displayed as arrows that are either filled or dashed, dashed arrows indicate the use of zero-padding and max pooling to match dimensions.

## 2.5 Network training

The proposed deep neural network was optimized iteratively using an average of the cross entropy cost for arousal and wake as

$$C(\theta) = -\frac{1}{N}\sum_{1}^{N} y_n \log(\hat{y}_n) + (1 - y_n)\log(1 - \hat{y}_n) \tag{1}$$

$$C_{total}(\theta) = \frac{1}{2}C_{arousal}(\theta) + \frac{1}{2}C_{wake}(\theta) \tag{2}$$

where $C(\theta)$ is the cost function with variables $\theta$, $N$ is the mini-batch size, $y_n$is the $n^{th}$ true label, and $\hat{y}_n$ is the $n^{th}$ predicted class probability. The dependency of the cost function $C$ on variables $\theta$ is given by the implicit dependency of $\hat{y}_n$ on $\theta$. The cross entropy cost $C_{total}(\theta)$ was minimized using the Adam optimization algorithm [40] with $\beta_1 = 0.9$, and $\beta_2 = 0.999$. Data was processed using a mini-batch size of $N = 300$ of 1 second bins, corresponding to 5 minutes of PSG. This ensures enough temporal context for prediction of arousals and wake. A learning rate $\eta = 10^{-3}$ was used and weights were initialized according to Glorot and Bengio (2010) [41].

The network was validated on a subset of 20 PSGs every 5,000 iterations, which was used to visually determine when network performance saturated. The network was trained for 350,000 iterations, corresponding to a total of 3,240 full night's PSGs. The architecture of the model was selected based on a hyper-parameter grid search, in which different number of LSTM cells and convolutional layers were used. In this grid search, the forward and backward LSTM layer had either 64, 128 or 256 cells, and a total of 13 or 19 convolutional layers. The optimal configuration of hyper-parameters was selected based on the minimal cross entropy cost obtained after 50,000 training iterations.

## 2.6 Probability Postprocessing

The cost function $C$ provides a measure of prediction error, however it does not directly provide information about how many arousal events are detected correctly. In fact, the model does not even predict events, but rather a probability of arousal or wake for each second. Events can be detected from the output probability by the proposed postprocessing, which binarizes the probability.

The postprocessing for the arousal probability P(Arousal) was implemented as follows

1) Threshold arousal probability, P(Arousal) > $T_{ar}$.
2) Connect arousal events closer than 10 seconds.
3) Discard detected events shorter than 3 seconds.

where the threshold $T_{ar}$ was optimized using the validation set. Rule 2 and 3 are based on the AASM manual [4] as scored arousals require a preceding 10 seconds of sleep and a duration of 3 seconds or more.

The postprocessing for the wake probability was implemented as

1) Threshold wake probability, P(Wake) > $T_w$.
2) Connect wake periods closer than 15 seconds.
3) Remove wake periods with duration less than 15 seconds.

where the wake threshold $T_w$ was also optimized using the validation set. Periods of wake are connected and discarded based on a 15 seconds criterion due to the gold standard, which states that 30-second epochs should be labelled as wake if more than half has the characteristics of wake (see introduction).

### 2.7 Model Validation and Testing of the Multimodal Arousal Detector

The mini-batch window processes 300 seconds of data at a time. During model validation and testing, the mini-batch window was moved 150 seconds at a time while using the central 150 seconds of each mini-batch, thereby excluding the ends of each window that do not have sufficient temporal context for optimal classification. The following set of performance metrics were used to evaluate the model. Predictions of arousals and wake in 1 second bins were compared to target labels. Detected arousal events do not have to match the labels to indicate the same event, therefore an arousal event true positive (TP) is defined as a predicted and target event having any overlap. In the context of measuring arousal scoring performance, true negatives (TNs) are trivial as predicting *non-arousal* is a very easy task. Performance of arousal scoring was therefore measured by precision and recall, defined as

$$\text{Precision} = \frac{TP}{TP + FP}, \quad \text{Recall} = \frac{TP}{TP + FN} \tag{3}$$

where TPs are either matching true 1 second bins or overlapping arousal events, FP are false positives, and FN false negatives. As both metrics are essential to measure performance, the F1 score was also used to summarize performance:

$$\text{F1} = 2 \cdot \frac{\text{Precision} \cdot \text{Recall}}{\text{Precision} + \text{Recall}} \tag{4}$$

In addition to F1 scores, the false positive rate (FPR), implemented as described in Eq. 5, was used to compare arousal scoring performance in different sleep stages.

$$FPR = \frac{FP}{FP + TN} \tag{5}$$

The FPR measure uses TN instead of TP to enable measuring performance in wake, where there are no TPs as arousals can only have onset during sleep by definition. The FPR metric for arousal events is defined as

$$FPR_{event} = \frac{FP_{event}}{FP + TN} \tag{6}$$

where the *event* subscript refers to number of events rather than 1 second bins.

To measure accuracy of classifying wake, TNs are important as they measure correctly identified sleep. Wake scoring performance is measured by recall, specificity and accuracy. The performance metrics of specificity and accuracy are defined by Eq. 7 and 8.

$$Specificity = \frac{TN}{TN + FP} \tag{7}$$

$$Accuracy = \frac{TP + TN}{TP + FP + FN + TN} \tag{8}$$

The probability postprocessing employ a set of thresholds $(T_{ar}, T_w)$ to binarize these probabilities. A threshold of 0.5 intuitively makes sense, as it distinguished between which class is more likely based on a cross entropy cost. However, this may not be optimal since performance is measured mostly by F1 score and events are connected and discarded during the postprocessing step. For this reason, the two thresholds, $T_{ar}$ and $T_w$, were calculated using the validation set by maximizing arousal event F1 score and wake accuracy, respectively.

As described above, a test set of 996 PSGs derived of multiple cohorts was used to calculate an unbiased estimate of MAD performance. We also used these data to evaluate performance in the cohorts and across sleep stages. This analysis gives insight as to if the model has any problems processing specific parts of the dataset. A second test set of 30 PSGs from each of the WSC and SSC datasets was used to compare performance of the model against that of multiple individual human scorers. PSG studies in this dataset had been scored five times using a pool of nine human scorers. In this analysis, individual scorer and model performance were evaluated based on a pseudo-consensus scoring as gold standard, which was established based on a majority vote of the four remaining scorers in a leave-one-out scheme. This was implemented by iterating through the five scorers for each PSG, creating a majority vote from the remaining four and calculating F1 score for both the scorer and model based on the consensus. The majority vote was defined as at least two scorers agreeing on an event i.e. an agreement of 50 % or more.

### 2.8 Relationship of Daytime Sleepiness with PSG arousals and wake events as detected using MAD

The MSL on an MSLT is considered a gold standard to measure daytime sleepiness objectively. We used a total of 1,447 PSGs with associated MSLT from the WSC to analyze the effect of sleep disruption as measured with the MAD on daytime sleepiness. The 1,447 PSG-MSLTs were performed in 873 subjects and Generalized Estimating Equations (GEE) statistics were used to account for repeated samples (see below).

To address properly the effects of arousals of all durations (i.e. detection of microarousals less than 15 sec and more than 15 sec leading a full-blown wake epoch), arousals and wake scored by our detector were combined to measure all sleep disruption and wakefulness. This global arousal measure of any transition to wake was computed as the union of predicted arousal and wake. As arousals are defined by the AASM to be preceded by at least 10 seconds of stable sleep, the new arousal measure was further postprocessed to combine wakefulness events preceded by fewer than 10 seconds of sleep.

We next examined the effect of arousals in the context of SDB and leg movements (LMs) occurrences. The WSC is a thoroughly investigated cohort, and both manual scoring and various prior event detectors have been proposed to score SDB events, blood oxygen desaturations and LMs, with high correlations with manually scored events. For SDB events, we applied the algorithm described by Koch et al. (2017) [42], which is a rule-based algorithm adapted to model the AASM 2012 manual [4] that detects 10 sec breathing disturbances with and without hypoxia events independently of any detected arousal (Breathing Disturbance Index, BDI, composed of BDI-Hypoxia and BDI non Hypoxia). LMs were scored using the Stanford PLM automatic detector (S-PLMAD) proposed by Moore et al. (2014) [29], which is also a rule-based algorithm adapted to the AASM scoring guidelines [4]. The S-PLMAD uses adaptive filtering of ECG artifacts and additional rules to account for specific noise types., and scores PLM independently of the presence of arousals. For more information about these detectors see Moore et al. and Koch et al. [42, 29].

#### 2.8.1 Coupling of automatically detected arousals to disturbed breathing or leg movement events

The distribution of new onset arousal/wake occurrences relative to SDB or LM events were investigated. Coupling rules were initially based on the AASM manual but revised if data suggested improvements could be made.

As illustrated in Fig. 4, and reported in Moore et al. [29] and others [43, 44], we found that the original AASM scoring definition which suggested disregarding LM within 0.5 sec of SDB events was not adequate as movements typically peaks a few seconds after the end of an SDB event. LMs secondary to SBDs were discarded if the onset of a LM occurred 15 seconds preceding a BD or -5 to 10 seconds at the offset of a SBD event. PLMs were then scored in agreement with the AASM manual, as at least four LMs not secondary to BDs with an inter-movement-interval between 5 and 90 seconds [5].

Breathing disturbances were coupled with arousals and desaturations based on similar distributions of time-locked events. The AASM manual do not state any objective rules for coupling these events, therefore the rules are based on the distributions visualized in Fig. 5 **(a)** and **(b)**. Arousals were coupled with BDs if the arousal onset in the interval of -5 to 10 seconds at the BD offset. Desaturations were coupled with BDS if the peak desaturation was between 5 and 35 seconds after the BD offset. PLMs were also coupled with arousals if their onsets were within a range of 5 seconds. The AASM manual uses a rule of minimum 0.5 seconds overlap of these events, but this rule is based on the distribution seen in Fig. 5 **(c)**.

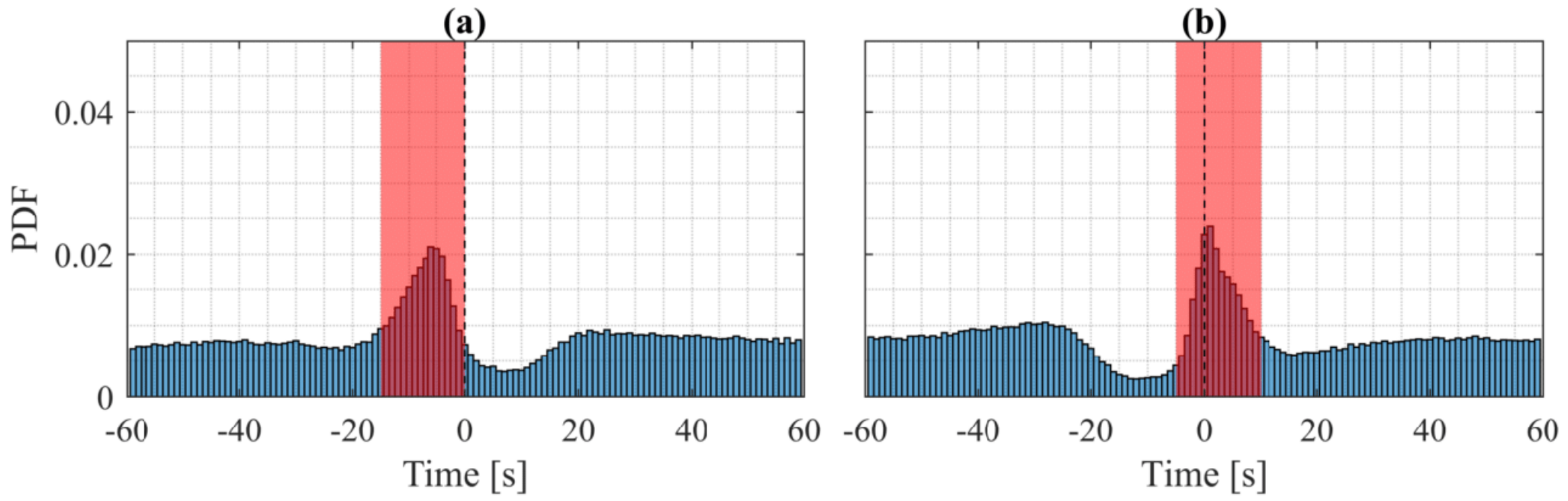

**Figure 4**: Discarding of LMs secondary to BDs based on the windows shown in red. **(a)** Distribution of LMs time-locked to the onset of BDs with no preceding BDs for 60 seconds. **(b)** Distribution of LMs time-locked to the offset of BDs.

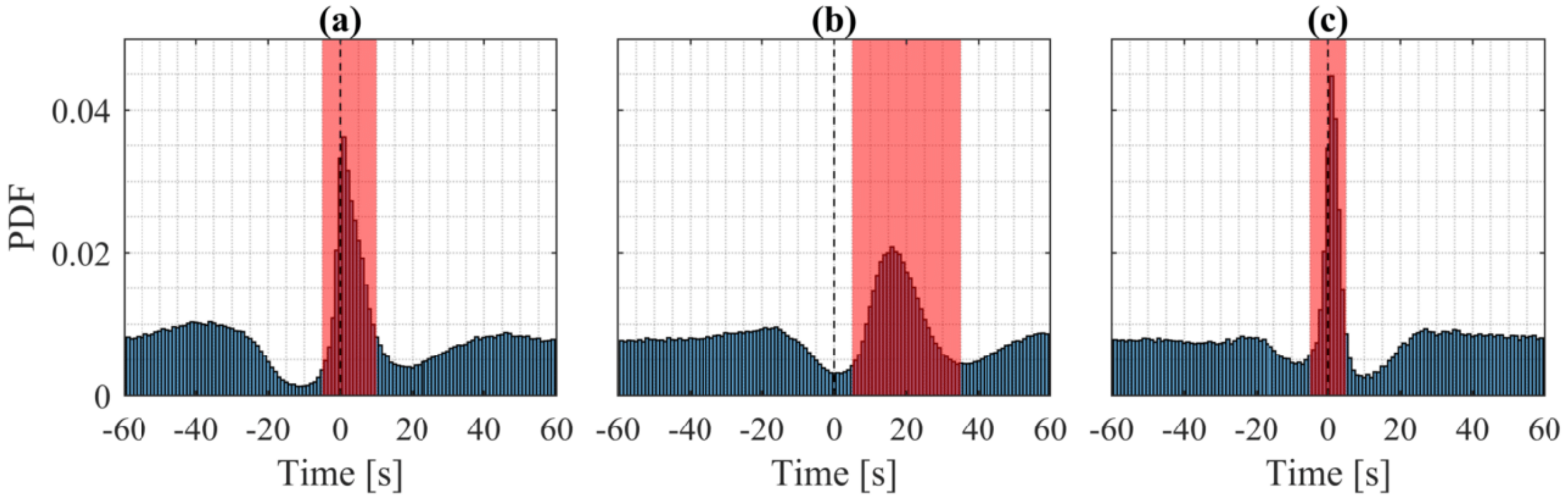

**Figure 5**: Event coupling based on relative distribution of time-locked events. **(a)** Distribution of peak desaturation time-locked to BD offsets. **(b)** Distribution of arousal onset time-locked to BD offsets. **(c)** Distribution of arousal onset time-locked to PLMs onset.

Based on these coupling rules, a set of PSG biomarkers was computed, which are described in Table 2. These were then used as covariates to explore how they best predict daytime sleepiness using the MSLT. In these analyses all variables were $log_2$-transformed to ensure normal distribution.

| PSG Biomarkers | | |
|---|---|---|
| **Name** | **Description** | **Mean ± Std** |
| BDI | BDs/h | 26.1 ± 14.1 |
| A-BDI | (BDs w/ arousal)/h | 7.6 ± 7.6 |
| NA-BDI | (BDs w/o arousal)/h | 18.1 ± 9.1 |
| H-BDI | (BDs w/ hypoxia)/h | 11.3 ± 12.1 |
| NH-BDI | (BDs w/o hypoxia)/h | 14.8 ± 6.8 |
| A-H-BDI | (BDs w/ arousal and hypoxia)/h | 4.9 ± 6.8 |
| A-NH-BDI | (BDs w/ arousal and w/o hypoxia)/h | 2.7 ± 2.3 |
| NA-H-BDI | (BDs w/o arousal and w/ hypoxia)/h | 6.2 ± 6.5 |
| NA-NH-BDI | (BDs w/o arousal and hypoxia)/h | 11.9 ± 5.4 |
| PLMI | PLMs/h | 11.4 ± 18.5 |
| A-PLMI | (PLMs w/ arousal)/h | 2 ± 3.3 |
| NA-PLMI | (PLMs w/o arousal)/h | 9.4 ± 16.5 |
| ArI | Arousals/h | 21.9 ± 10 |
| Spon-ArI | (Spontaneous arousal)/h | 12.4 ± 4.9 |
| Me-A-Dur | Median arousal duration (s) | 9.5 ± 1.5 |
| Me-BD-A-Dur | Median BD arousal duration (s) | 10.4 ± 5 |
| Me-PLM-A-Dur | Median PLM arousal duration (s) | 12.3 ± 33.9 |
| Me-Spon-A-Dur duration | Median spontaneous arousal (s) | 9.3 ± 2.3 |
| TST | Total Sleep Time (h) | 6.7 ± 1 |
| WASO | Wake after sleep onset (h) | 1 ± 0.7 |

**Table 2**: Description of PSG biomarkers. BD: breathing disturbance, PLM: periodic leg movement.

### 2.8.2 Statistical Analysis

We first explored collinearity of the $\log_2$transformed variables described in Table 2 using Pearson correlation coefficient statistics. Next, the effect of each variable on MSLT was examined using generalized estimating equations (GEE), a statistical technique that maximizes power for cross-sectional analyses that have repeated measures in some subjects [45] to estimate robust standard errors and allow for usage of repeated measurements. GEE was implemented in GEEQBOX [46], a MATLAB toolbox for GEE, using the identity link function and the first order autoregressive correlation structure. This correlation structure is well suited for repeated measurements that are roughly evenly spaced in time [46], which is the case for the WSC as repeated measurements for subjects are conducted as 4-year follow ups. The *sandwich* *estimator* [45] was used to provide a robust estimate of the standard errors. Finally, stepwise linear regression was implemented to find an optimal set of variables that provided most information about the MSL. For stepwise linear regression a subset of 873 PSGs was used, as repeated measurements could not be included. The best regression model was determined based on the adjusted coefficient of determination $R^2$, which is given by

$$\bar{R}^2 = 1 - \frac{SS_{res}/(n-p-1)}{SS_{tot}/(n-1)} \tag{9}$$

where $SS_{res}$ is the sum of squares of the residuals, $SS_{tot}$ is the total sum of squares, $n$ is the number of observations, and $p$ is the number of explanatory variables. When additional explanatory variables are added to a model, the adjusted $R^2$ will only increase if $R^2$ increases by more than expected by chance. Of note, because many variables are colinear, the set of variables selected by the model may not necessarily represent the ideal best variable if a larger dataset was analyzed. It nonetheless helped us validate key features of the detector.

*Code availability:*

The Matlab and Python code for the MAD detector is available on GitHub at: https://github.com/Stanford-STAGES/multimodal-arousal-detector.

## 3 Results

### 3.1 Network Performance

### 3.1.1 Probability Threshold

The effect of changing probability thresholds for arousal and wake detection was examined by varying threshold in steps of 0.025. For each threshold, precision, recall, and F1 score were calculated for arousal events while sensitivity, recall, and accuracy were calculated for the wake predictions. Arousal predictions in 1 second bins will from this point referred to as *arousal samples* for convenience. The thresholds that maximize arousal event F1 score and wake accuracy were selected, which are $T_{ar} = 0.225$ and $T_w = 0.45$. The precision-recall (PR) curves for arousal predictions is shown in Fig. 6 and the receiver operating characteristic (ROC) curve for wake are shown in Fig. 7.

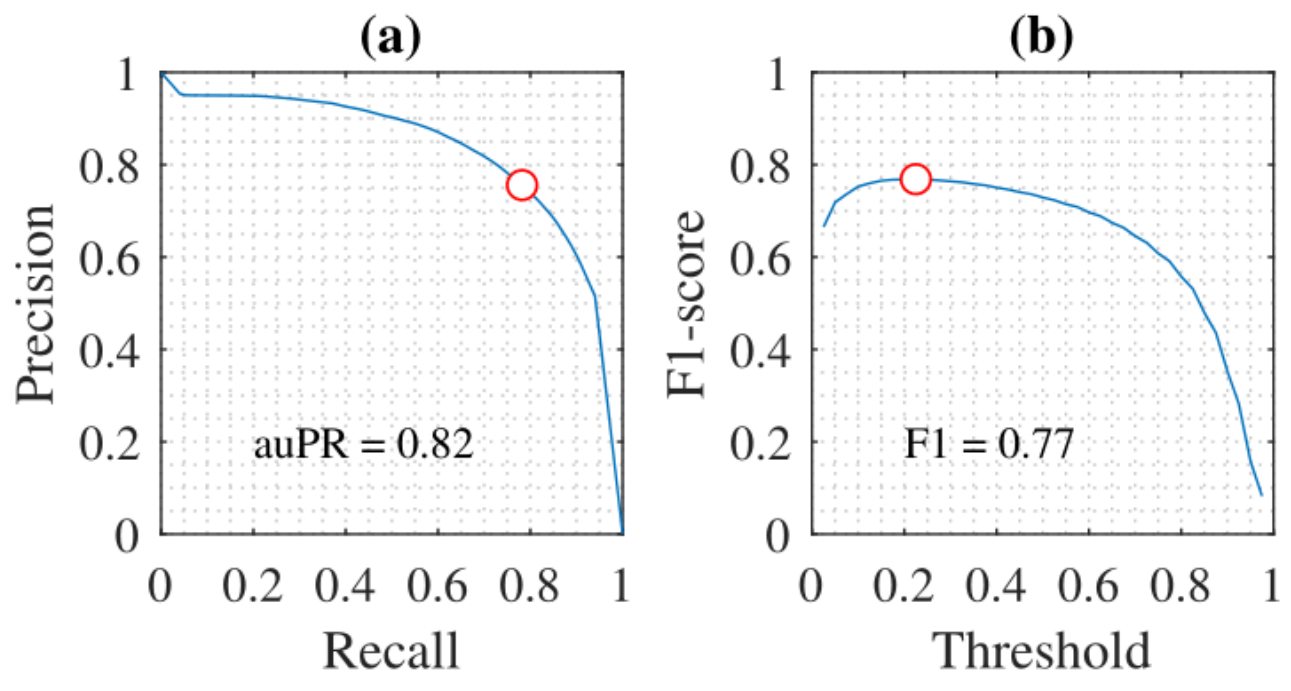

**Figure 6**: PR-curves and ROC-curves for predictions. The red circle indicates the optimal threshold at $T_{ar} = 0.255$. **(a)** PR-curve for arousal events. **(b)** F1 score for arousal events.

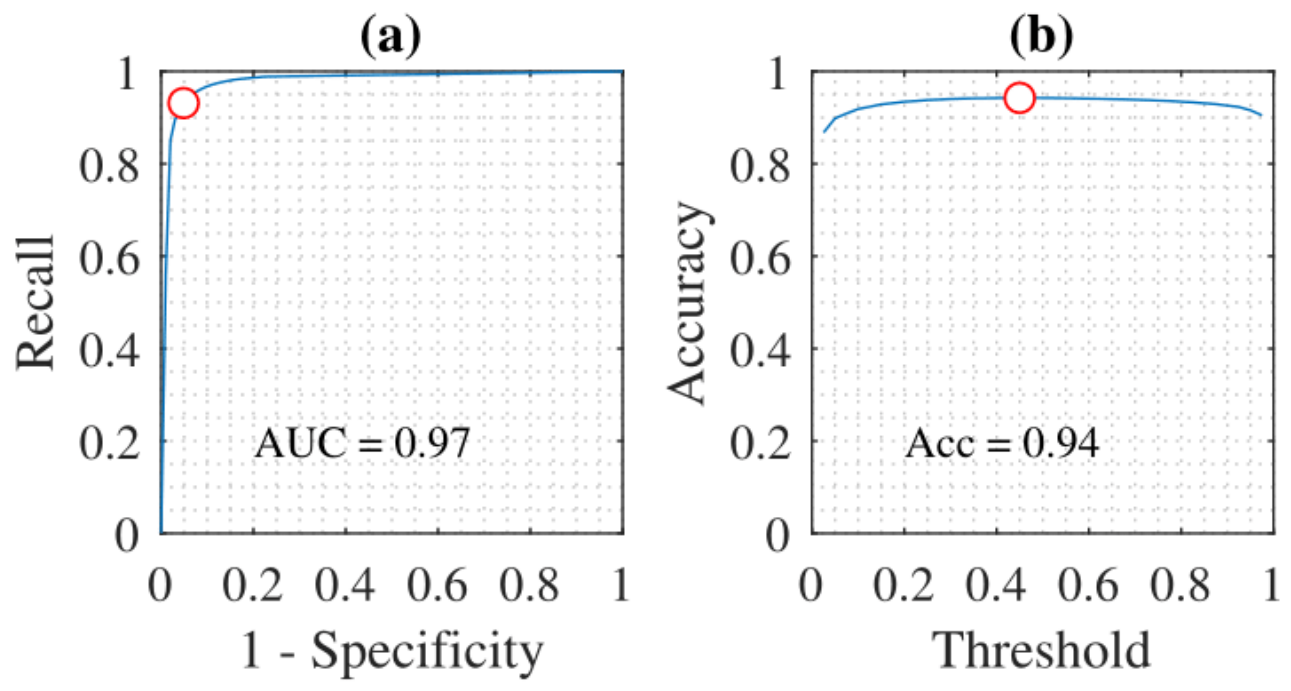

**Figure 7**: ROC-curves for predictions. The red circle indicates the optimal threshold at $T_w = 0.45$. **(a)** ROC-curve for wake. **(b)** Accuracy for wake.

### 3.1.2 Test Performance

The model was evaluated on the test set of 996 PSGs from MrOS, CFS, and WSC. Predictions were postprocessed using the optimal thresholds $T_{ar} = 0.225$ and $T_w = 0.45$. Fig. 8 show an example of arousal and wake predictions over a full night's PSG, and Fig. 9 shows predictions of a segment of the same PSG in a 60 seconds window at an arousal prediction.

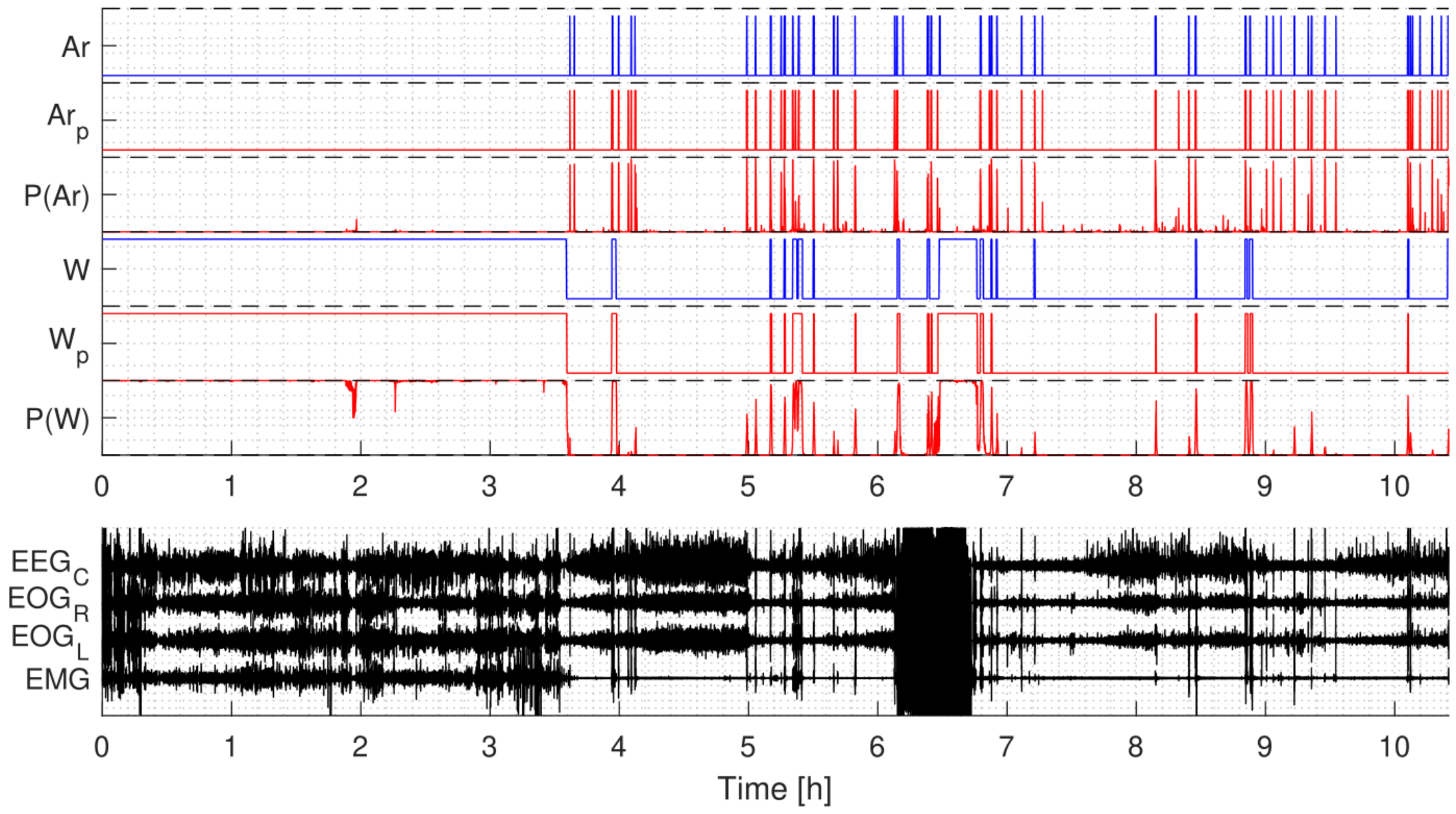

**Figure 8**: Example of arousal and wake predictions for a full night's PSG. P(Ar) and P(W) are the probability output, $Ar_p$ and $W_p$ are the predicted labels, while Ar and W are the target labels. The predictions for both arousals and wake are in agreement with the target labels.

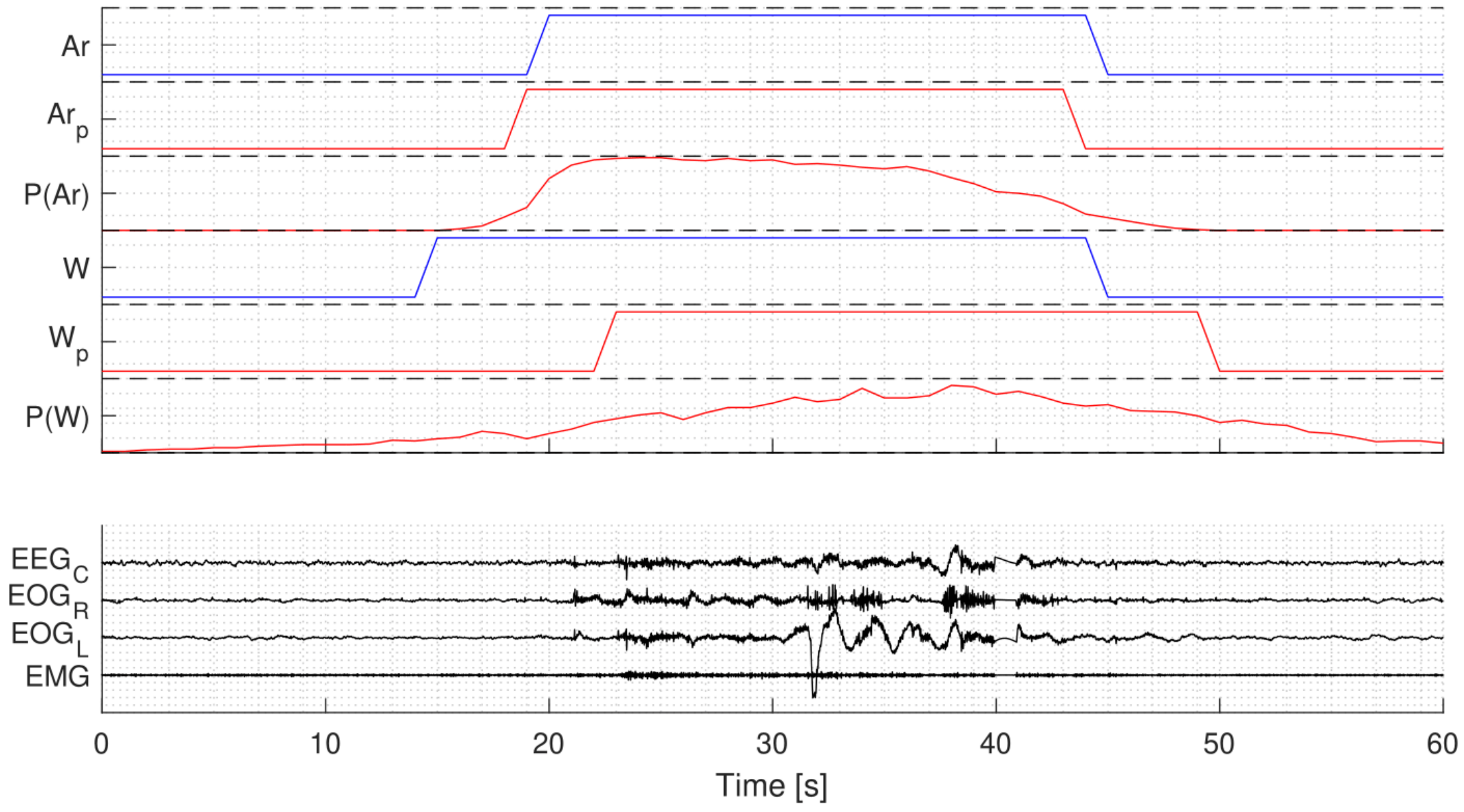

**Figure 9**: Example of arousal and wake predictions in a 60 seconds segment from the same PSG displayed in Fig. 8. P(Ar) and P(W) are the probability output, $Ar_p$ and $W_p$ are the predicted labels, while Ar and W are the target labels.

Table 3 shows the classification performance on the different cohorts as well as all test data. Arousal events were detected with an F1 score of 0.76 over all test data. On the WSC data, arousal events were detected with a lower F1 score of 0.70. The slightly performance in the WSC is likely caused by two factors: the fact that this data was unseen during training and the fact arousals in this dataset were only annotated as onsets, hence all annotated arousals were assumed to have a duration of 3 seconds, as this is the minimum duration of AASM arousals. Wake was predicted with an accuracy of 0.95 across all test data. The wake prediction accuracy was 0.93 on the WSC data, which suggests that the model works almost as well on data from sources unseen during training.

| | **Arousal Event** | | | **Arousal Samples** | | | **Wake** | | |
|---|---|---|---|---|---|---|---|---|---|
| | **Precision** | **Recall** | **F1** | **Precision** | **Recall** | **F1** | **Specificity** | **Recall** | **Accuracy** |
| **MrOS** | 0.77 | 0.81 | 0.79 | 0.65 | 0.73 | 0.69 | 0.96 | 0.93 | 0.95 |
| **CFS** | 0.69 | 0.83 | 0.75 | 0.61 | 0.76 | 0.68 | 0.97 | 0.92 | 0.95 |
| **WSC** | 0.62 | 0.82 | 0.7 | - | - | - | 0.97 | 0.84 | 0.93 |
| **All** | 0.72 | 0.81 | 0.76 | 0.64 | 0.73 | 0.68 | 0.96 | 0.92 | 0.95 |

**Table 3**: Classification performance on test set.

Variation in model's arousal event scoring performance is displayed in Fig. 10 as a scatter plot, which shows a good performance for the clear majority of PSG recordings, but also a set of PSGs with poor performance In Fig. 10, the size of each dots is proportional to the number of arousals detected in a single PSG, so that small dots that have poor performance could reflect limited sample size for performance evaluation. Through visual inspection it was observed that there was PSGs with mostly missing arousal target labels or target labels that appears random. This suggests that a substantial part of the error on PSG recordings with very poor performance is caused by human error.

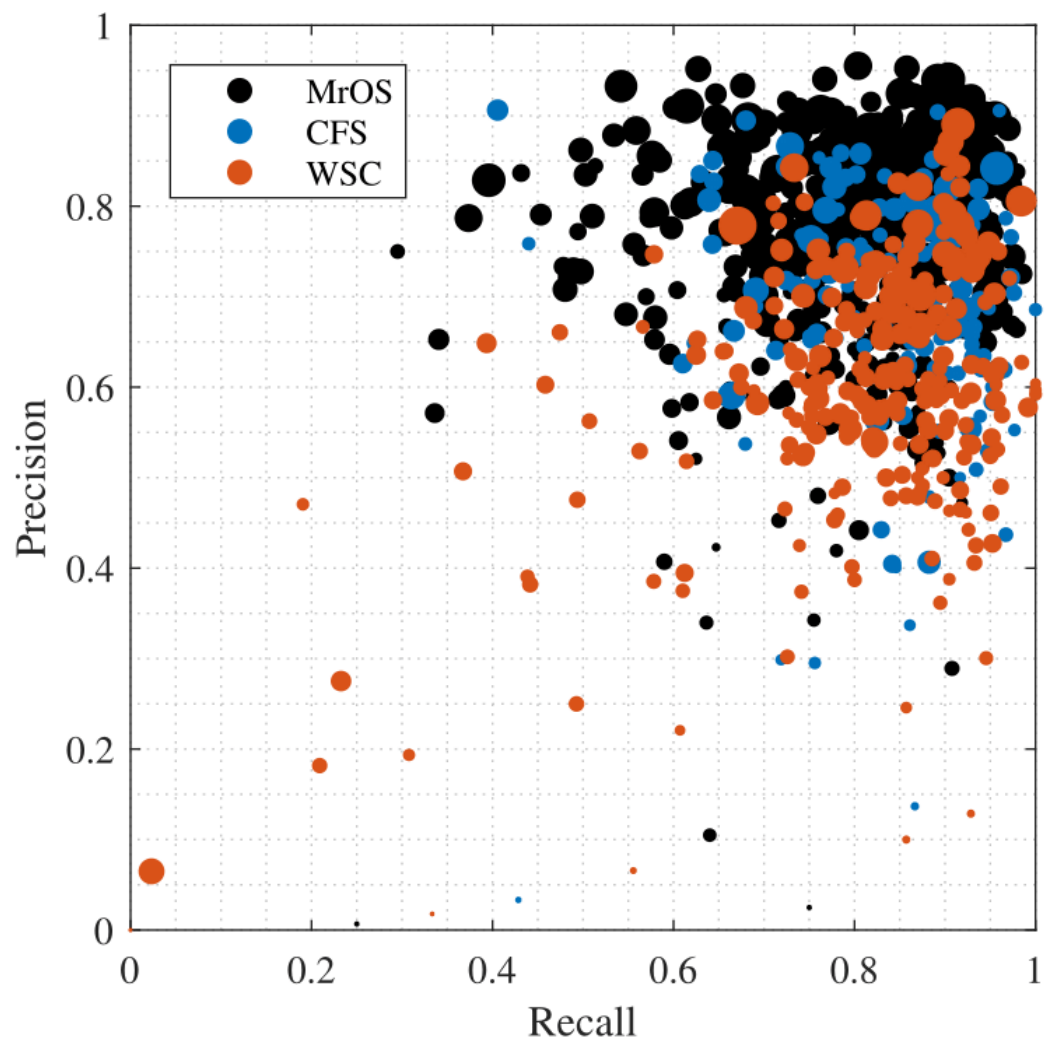

**Figure 10**: Precision-recall scatter plot of all test data. Each dot represents a full PSG, and the size is proportional to the number of annotated arousal events, ranging from 1 to 498.

Human scoring bias was further investigated by computing the average arousal annotation frequency for each 30-second epoch. Arousals and their definition according to the AASM guidelines [4] is completely unrelated to the 30-second epoch. However, as shown in Fig. 11, arousal annotations in the MrOS and CFS cohort show a clear bias toward the central and late part of the 30-second epochs. This bias is likely to have been introduced as a result of software that visualize data in 30-second epochs.

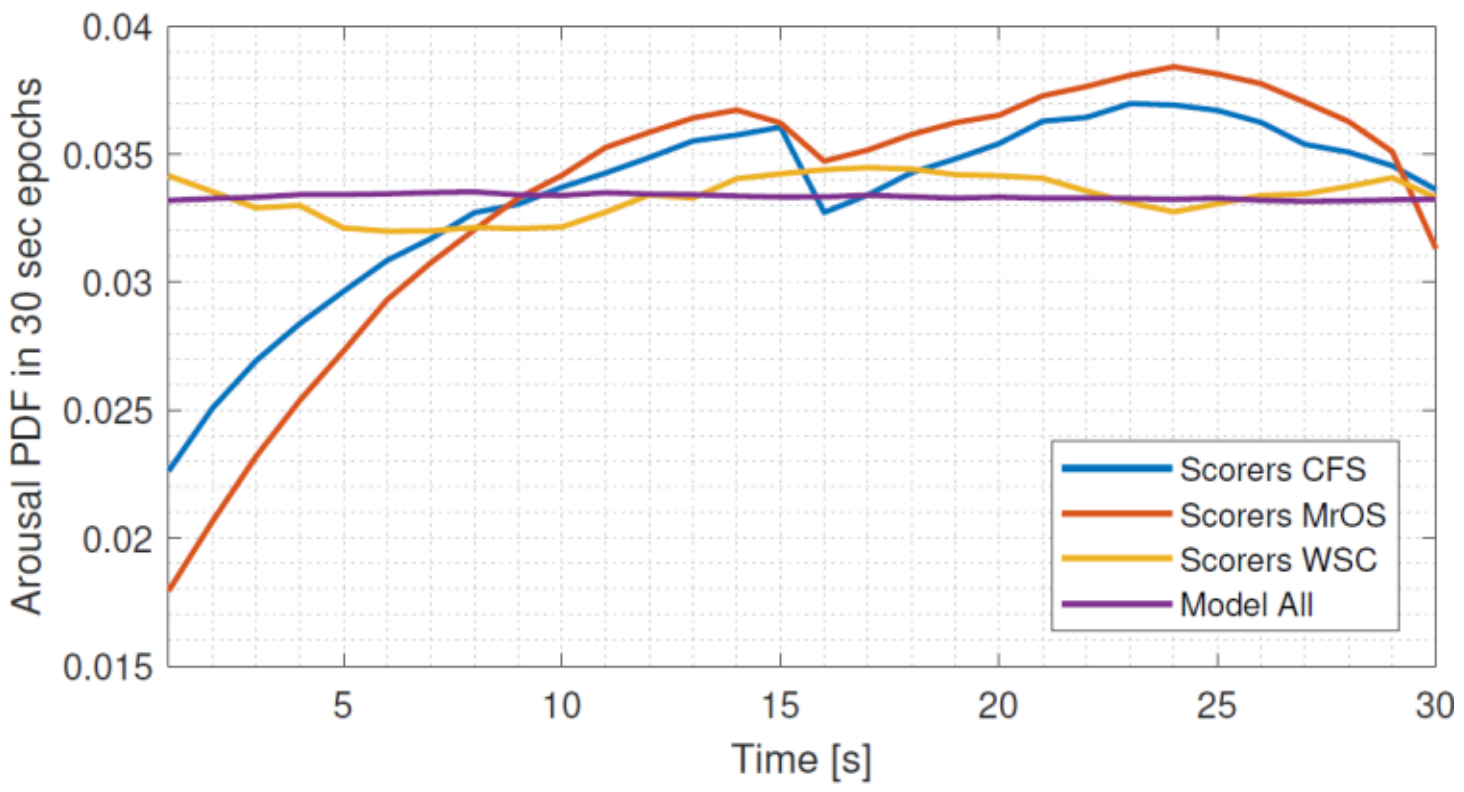

**Figure 11**: Arousal PDF over 30-second epochs for manual annotations in CFS (*n* = 145), MrOS (*n* = 580), WSC (*n* = 271) and model predictions over all test data (*n* = 996). The distribution is expected to be uniform as scoring of arousals is unrelated to this epoch, however annotations in CFS and MrOS exhibit a strong bias toward the central and later part of the 30-second epoch.

The dependency of the model's classification performance on sleep stage has also been examined. Table 4 and 5 shows the performance in the different sleep stages for arousals and wake/sleep, respectively.

| Arousal | | | | | | |
|---|---|---|---|---|---|---|
| | | Wake | N1 | N2 | N3 | REM |
| **Events** | FPR | 6.1e-4 | 2.5e-3 | 1.8e-3 | 6.2e-4 | 1.6e-3 |
| | Precision | - | 0.79 | 0.77 | 0.69 | 0.74 |
| | Recall | - | 0.64 | 0.83 | 0.86 | 0.86 |
| | F1 | - | 0.71 | 0.8 | 0.77 | 0.8 |
| **Samples** | FPR | 0.016 | 0.026 | 0.026 | 0.007 | 0.022 |
| | Precision | 0.64 | 0.7 | 0.66 | 0.56 | 0.6 |
| | Recall | 0.75 | 0.5 | 0.75 | 0.76 | 0.76 |
| | F1 | 0.69 | 0.58 | 0.7 | 0.65 | 0.67 |

**Table 4**: Arousal Scoring performance in the different sleep stages.

Table 4 shows that arousal events are detected well in sleep, although with a relatively poorer performance in N1. The FPR for arousal events is the lowest in wake and highest in N1. The performance metrics for the arousal samples also show a good performance in all stages, but with a slight decrease in performance in N1.

| Wake/Sleep Accuracy | | | | | |
|---|---|---|---|---|---|
| | **Wake** | **N1** | **N2** | **N3** | **REM** |
| **MrOS** | 0.935 | 0.746 | 0.977 | 0.996 | 0.965 |
| **CFS** | 0.916 | 0.805 | 0.976 | 0.996 | 0.989 |
| **WSC** | 0.842 | 0.792 | 0.981 | 0.997 | 0.99 |
| **All** | 0.919 | 0.771 | 0.978 | 0.996 | 0.975 |

**Table 5**: Wake and sleep classification performance in the different sleep stages.

The wake/sleep accuracy displayed in Table 5, showing that sleep is detected well in N2, N3, and REM, while accuracy is lower in N1.

### 3.1.3 Comparison to Multiple Scorers

The performance of the model was compared to multiple scorers on a dataset of 60 PSGs. The model and human scorer predictions were evaluated with respect to a pseudo-consensus of multiple scorers. The comparison was based on the arousal event F1 score. The results of this test are presented in Table 6.

| F1 score | | A | B | C | D | E | F | G | H | I | Mean | Mean (A - H) |
|---|---|---|---|---|---|---|---|---|---|---|---|---|
| Human Scorer | μ | 0.62 | 0.57 | 0.65 | 0.68 | 0.62 | 0.65 | **0.71** | 0.61 | 0.32 | 0.60 | 0.64 |
| | (σ) | (0.17) | (0.16) | (0.14) | (0.16) | (0.19) | (0.14) | (0.11) | (0.17) | (0.2) | (0.19) | (0.16) |
| Model | μ | **0.7** | **0.67** | **0.68** | **0.69** | **0.70** | **0.72** | 0.7 | **0.66** | **0.71** | **0.69** | **0.69** |
| | (σ) | (0.14) | (0.14) | (0.11) | (0.1) | (0.13) | (0.08) | (0.08) | (0.12) | (0.12) | (0.12) | (0.11) |
| *p*-val | | **0.033** | **0.006** | 0.318 | 0.755 | **0.038** | **0.016** | 0.67 | 0.19 | **1.6e-14** | **6.9e-12** | **1.7e-5** |

**Table 6**: Arousal event F1 score of human scorers and model on pseudo-consensus. *p*-values below 0.05 are highlighted as significant and are calculated using a two-sample t-test with the null hypothesis being equal means while assuming equal but unknown variances.

As seen in Table 5, the model predicts arousals with a significantly higher average F1 score in comparison to 5 scorers, while there is no significant difference to the remaining 4 scorers. Further, the model outperforms the average scorer with a difference in F1 score of 0.09. Scorer I performed particularly bad in comparison, however the model also outperforms scorer A – H on average. This indicates that our automatic arousal scoring system performs substantially better than human scorers.

**3.2 Statistical Analysis of Daytime Sleepiness**

The statistical analysis was performed using a combined measure of arousal and wake, which does not discriminate based on the 15 seconds threshold used in current scoring rules [4]. The duration of predicted arousals has the distribution shown in Fig. 12. The distribution peaks with an arousal duration of 9 seconds and decreases exponentially onward. The data does not provide any justification as to split the arousal measure at 15 seconds.

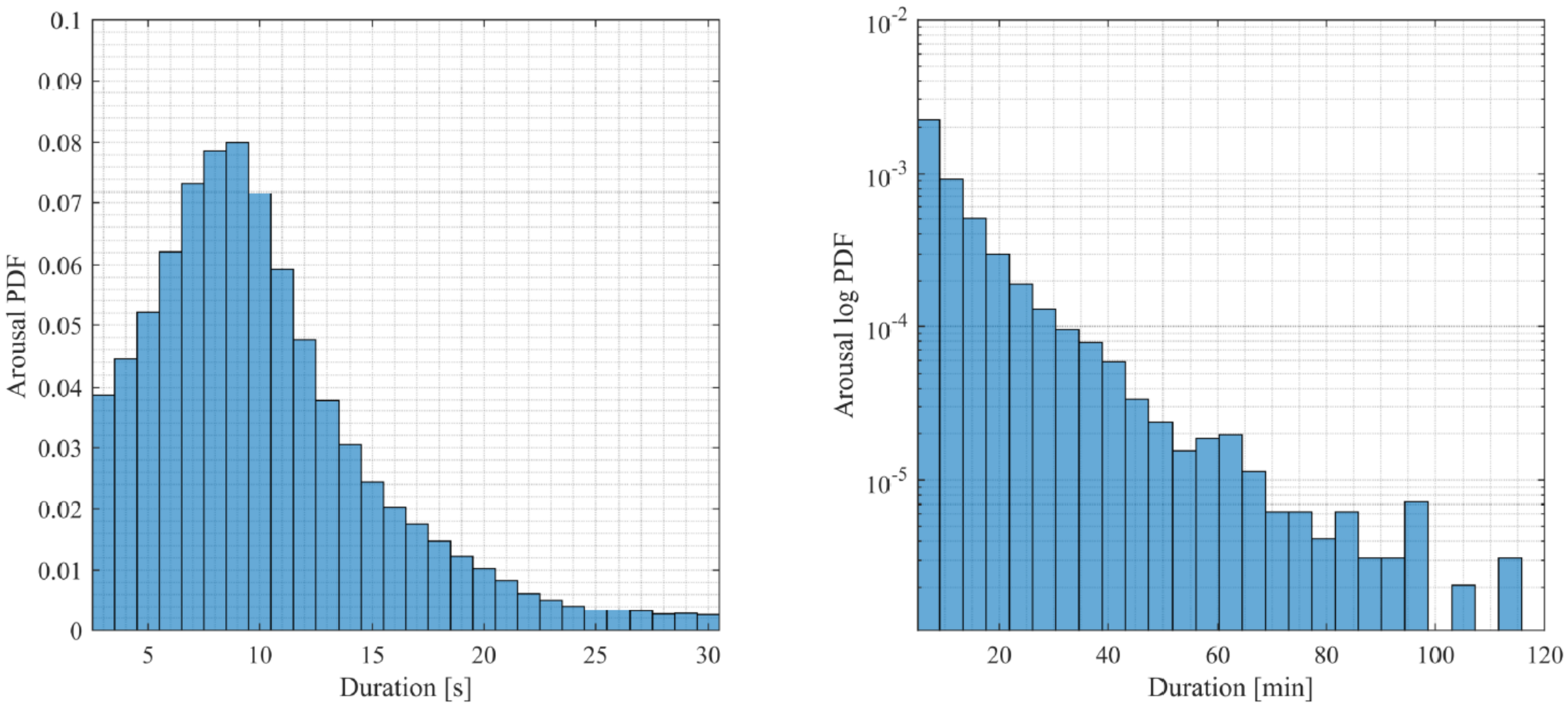

**Figure 12**: Distribution of duration of predicted arousals in WSC PSG data ($n$ = 1447). The left plot shows the distribution peaking at 9 seconds, while the right figure shows that the probability decreases exponentially as a function of duration in minutes (shown as a linear decrease in a logarithmic plot).

Fig. 13 shows a correlation matrix of $\log_2$-transformed sleep variables, computed using Pearson correlation coefficient statistics.

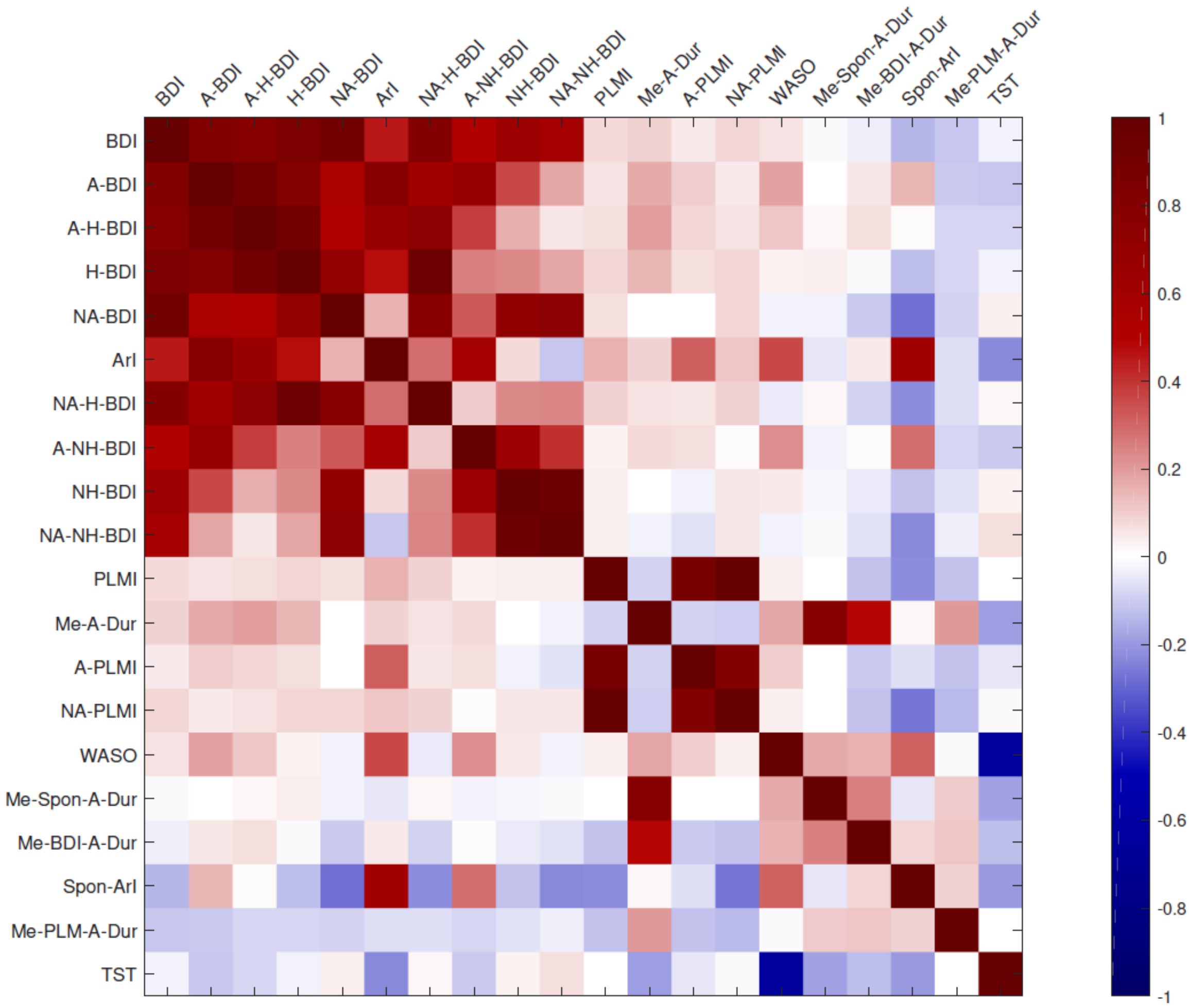

**Figure 13**: Correlation of biomarker variables after $\log_2$-transform. Correlation are sorted in a descending order from left to right. The description of these biomarkers can be found in Table 2.

Results show that the BDI is highly correlated to all variables related to breathing disturbances, while the other breathing disturbance variables are correlated in a more complex pattern. As reported in Kock et al. [42], breathing disturbances associated with hypoxia are only weakly correlated with those not associated with hypoxia, thus two clusters are apparent. Arousal-associated SDB (A-BDI) encompass almost all SDB events, which indicate that our previously reported non-hypoxia-BDI (NH-BDI) really corresponds to SDB events associated with arousals but no desaturation (A-NH-BDI). The correlation between A-H-BDI and NA-NH-BDI was shown to be insignificant ($r$ = 0.052, $p$ = 0.13), suggesting that subjects can be affected by either subtype of sleep disordered breathing.

The sleep variables relating to PLMs were also all highly correlated. The Me-Spon-A-Dur variable has no correlation to either breathing disturbance variables, PLMI or the ArI.

A GEE was used to estimate model parameters for each sleep variables at a time as seen in Table 7. These estimates were adjusted for age, BMI, sex, habitual sleep duration, minutes of sleep the two nights preceding the MSLT, and predicted WASO. The predicted WASO was included as it and arousal metrics are positively correlated (see Fig. 13), while WASO and arousal metrics have an inverse relation to MSLT as displayed in Table 7. Due to this interaction it was necessary to add it to the linear models to show the effect of arousals.

| **MSL Model Parameters** | | | | | | |
|---|---|---|---|---|---|---|
| **Predictive Variables** | **β Coefficient** | **95 % CI** | ***p*-value** | **β Coefficient*** | **95 % CI*** | ***p*-value*** |
| WASO | - | - | - | 1.45 | (0.94, 1.96) | 3.1e-8 |
| BDI | -0.65 | (-0.97, -0.34) | 4.5e-5 | -0.66 | (-0.97, -0.34) | 4.6e-5 |
| A-NH-BDI | -0.63 | (-0.96, -0.31) | 0.00014 | -0.44 | (-0.77, -0.11) | 0.008 |
| NH-BDI | -0.61 | (-0.94, -0.28) | 0.00028 | -0.57 | (-0.90, -0.24) | 0.0008 |
| A-BDI | -0.46 | (-0.71, -0.21) | 0.00032 | -0.38 | (-0.64, -0.13) | 0.003 |
| NA-BDI | -0.51 | (-0.84, -0.18) | 0.0022 | -0.57 | (-0.90, -0.24) | 0.0007 |
| A-H-BDI | -0.34 | (-0.56, -0.12) | 0.0027 | -0.32 | (-0.54, -0.1) | 0.0046 |
| Me-Spon-A-Dur | -1.5 | (-2.4, -0.51) | 0.0028 | -0.95 | (-1.9, 0.004) | 0.051 |
| NA-NH-BDI | -0.49 | (-0.81, -0.16) | 0.0036 | -0.50 | (-0.83, -0.16) | 0.0034 |
| ArI | -0.67 | (-1.2, -0.18) | 0.0075 | -0.22 | (-0.7, 0.26) | 0.37 |
| H-BDI | -0.26 | (-0.46, -0.05) | 0.014 | -0.29 | (-0.5, -0.08) | 0.006 |
| TST | -1.2 | (-2.8, 0.32) | 0.12 | -2.94 | (-2.8, 0.32) | 5.3e-6 |
| NA-H-BDI | -0.18 | (-0.41, 0.05) | 0.13 | -0.26 | (-0.49, -0.02) | 0.03 |
| Me-PLM-A-Dur | 0.2 | (-0.13, 0.53) | 0.24 | 0.2 | (-0.13, 0.53) | 0.24 |
| NA-PLMI | -0.071 | (-0.2, 0.06) | 0.29 | -0.073 | (-0.21, 0.06) | 0.29 |
| Me-BDI-A-Dur | 0.31 | (-0.37, 1) | 0.37 | 0.53 | (-0.16, 1.22) | 0.13 |
| PLMI | -0.058 | (-0.19, 0.07) | 0.38 | -0.058 | (-0.19, 0.07) | 0.38 |
| Me-A-Dur | -0.44 | (-1.7, 0.79) | 0.49 | 0.13 | (-1.08, 1.34) | 0.84 |
| A-PLMI | -0.074 | (-0.31, 0.16) | 0.53 | -0.03 | (-0.27, 0.21) | 0.81 |
| Spon-ArI | 0.011 | (-0.51, 0.53) | 0.97 | 0.44 | (-0.06, 0.95) | 0.084 |

**Table 7**: MSL parameters describing the relationship between the $\log_2$transformed sleep variables and MSL. The table is sorted with respect to *p*-value in a descending order. The parameter estimates are adjusted for age, BMI, sex, habitual sleep duration, minutes of sleep the two nights preceding the MSLT, and predicted WASO. *The effect of the predictive variables is shown when WASO is not included. The effect of the adjusted parameters is shown in Table 8.

| MSL Model Adjusted Parameters | | | |
|---|---|---|---|
| **Predictive Variables** | **β Coefficient** | **95 % CI** | ***p*-value** |
| Age | 0.09 | (0.57,0.12) | 1e-7 |
| Sex = female | -1.35 | (-1.89, -0.82) | 7.6e-7 |
| Min night 0 | 0.0076 | (0.004, 0.011) | 1.5e-5 |
| HSD | 0.49 | (0.22, 0.76) | 3.4e-4 |
| Min night 1 | 0.0049 | (0.002, 0.008) | 0.001 |
| BMI | -0.06 | (-0.10, -0.02) | 0.0027 |
| Intercept | -1.37 | (-4.58, 1.85) | 0.40 |

**Table 8**: Effect of adjusted parameters on MSL. The table is sorted with respect to *p*-value in a descending order.

The results in Table 7 show that the ArI, Me-Spon-A-Dur, and most breathing disturbance variables have a significant ($p < 0.05$) negative effect on the MSL.

A series of MSL models was fitted with stepwise linear regression to provide information as to which sleep variables provides most independent information about the MSL. These models were fitted using the first available visit of each subject, which reduces the number of observations to 873. The results of these models are summarized in Table 9.

| # | Model | *p*-value | $R^2$ (adj) |
|---|---|---|---|
| 1 | - | 2.85e-25 | 0.134 |
| 2 | **BDI** | 2.43e-27 | 0.146 |
| 3 | **A-BDI**, NA-BDI | 1.21e-26 | 0.143 |
| 4 | **A-H-BDI**, NA-H-BDI | 3.62e-26 | 0.14 |
| 5 | **A-NH-BDI**, NA-NH-BDI | 1.82e-26 | 0.142 |
| 6 | A-H-BDI, **A-NH-BDI** | 1.82e-26 | 0.142 |
| 7 | **A-H-BDI**, A-NH-BDI, NA-H-BDI, **NA-NH-BDI** | 6.58e-27 | 0.146 |
| 8 | **NA-BDI**, **ArI** | 1.75e-26 | 0.144 |
| 9 | NA-H-BDI, **NA-NH-BDI**, **ArI** | 7.66e-27 | 0.146 |
| 10 | **NA-NH-BDI**, **ArI**, **Me-Spon-A-Dur** | 1.34e-27 | 0.151 |
| 11 | **BDI**, **Me-Spon-A-Dur** | 8.84e-28 | 0.15 |

**Table 9**: Variables included (shown in bold font) with stepwise linear regression in a set of fitted models. The models are adjusted for age, BMI, sex, habitual sleep duration, minutes of sleep the two nights preceding the MSLT, and predicted WASO. WASO had a significant effect in all models.

Model 1 in Table 9 shows the explanatory power of the variables that each model is adjusted for. Breathing disturbances associated with arousals seems to have a stronger effect on the MSL as indicated by model 2 to 6. The A-H-BDI and NA-NH-BDI were shown in model 7 to be independent measures of breathing disturbances that both affects the MSL. Model 8 and 9 shows that the ArI and NA-BDI have an independent effect on MSL. The results of model 10 and 11 show that the inclusion of Me-Spon-A-Dur further strengthens the MSL models. The model with the most explanatory power as measured by the adjusted $R^2$ included the sleep variables NA-NH-BDI, ArI, and Me-Spon-A-Dur. Surprisingly, PLMI-A was not associated with MSL in any model.

## 4 Discussion

### 4.1 Performance of MAD: Comparison to Multiple Scorers

The arousal event classification performance of our model was compared to that of nine individual scorers by evaluating predictions with respect to a pseudo-consensus. The pseudo-consensus was based on majority voting from the remaining four human scorers is not expected to be near perfect as the discrepancies between scorers is so large but is assumed to be good enough to justify the comparison to individual scorers. The comparison showed that the model, in terms of F1 score, significantly outperformed five of nine individual human scorers, while there was no significant difference to the remaining four. Further, the model outperformed the average human scorer with a difference in F1 score of 0.09.

The significance of this comparison is further emphasized by the fact that the performance of the model is unbiased, as the data from the WSC and SSC were unseen during training. The best performing human scorer achieved a F1 score of 0.71, which also suggests that the F1 score of 0.76 achieved on the larger dataset of 996 PSGs is satisfactory. In brief, our detector was able to score arousals better than most scorers.

### 4.2 Comparison to Previous Methods

Comparing the proposed arousal detecting system to existing published methods is difficult due to differences in used data, number of human scorers, scoring unit, performance metric etc. The system with highest reported performance was published by Sorensen et al. (2012) with an F1 score of 0.87 using cross-validation on 24 subjects. However, we believe the model would be unlikely to generalize on unseen data as it is trained on annotations from a single human scorer. Furthermore, this high performance suggest overfitting to this single scorer, as the best human scorer in this study only achieved a F1 score of 0.71. The general performance of the arousal scoring system proposed in this project is considered better, as it is fully automatic and was evaluated on a much larger dataset annotated by many human scorers. In general, the proposed method stands out from existing methods on the points of it being fully automatic and having a robust performance on a very large dataset.

Wake was predicted with an overall accuracy of 0.95 in 1 second bin (epochs), which is different than the standardized 30-second epoch. The sleep/wake accuracy of the model is high, but it is inflated to some degree due to long periods of wake between recording start time and sleep onset in MrOS and CFS data as shown in Fig. 8. Accuracy could have been measured between annotations of lights off/on, however wakefulness during time of lights on was also considered as wake. This also has the effect of simplifying the preprocessing and does not require manual annotations. The sleep/wake prediction performance of the model can be compared to the performance of published methods for sleep staging. The studies by Biswal et al. (2017) and Sun et al. (2017) both used a test set of 1000 PSGs similar to this project and achieved a wake/sleep accuracy of 0.937 and 0.929, respectively [33, 47]. It should be noted that the wake accuracy from these studies was estimated from normalized confusion matrices using the distribution of sleep stages of data from the WSC. Stephansen et al. (2017) and Supratak et al. (2017) achieved a higher wake/sleep accuracy of 0.967 and 0.961, respectively [3, 34]. The wake accuracies reported in these studies were measured using smaller test sets of 70 and 82 PSG. The wake accuracy of the proposed system is at the level of state-of-the-art sleep staging methods, but it is difficult to infer which method clearly perform best.

### 4.3 Statistical Analysis of Daytime Sleepiness in relation to arousal

Regression analysis revealed a significant association between the breathing disturbance sleep variables and the MSL on the MSLT, except for NA-H-BDI. The NA-H-BDI is expected to be highly noisy as most breathing disturbances with hypoxia are suspected to provoke an arousal. The effect of breathing disturbances on MSL has previously been demonstrated by multiple studies [42, 48, 49]. The stepwise linear regression model showed that breathing disturbances can be described with two independent variables A-H-BDI and NA-NH-BDI, which agrees with previous findings by Koch et al. (2017), who showed that H-BDI and NH-BDI are independent measure of sleep disordered breathing in the same dataset. The significance of the breathing disturbances without hypoxia or arousal is slightly counter-intuitive as the breathing disturbances do not provoke any measured physiological changes. A possible explanation is that these types of breathing disturbances may result or be a result of subcortical disturbances, which impair the restorative effect of sleep.

More surprisingly, sleep variables related to PLMs had no effect on MSL, even in the presence of an associated arousal. Chervin et al. (2001) found a weak, but significant association between PLMs and MSL [50]. In this sample, PLMs with and without arousals seem to either have a negligible or a small effect on MSL. It should be noted that a *p*-value above 0.05 simply implies a lack of evidence to reject the null hypothesis, rather than showing that the sleep variable has no effect on daytime sleepiness. The regression coefficient as well and standard error is also highly dependent on the variables that the model is adjusted for, the test therefore only provides information about the relation to MSL based on the knowledge of the variables adjusted for. Together with the finding described above suggesting that breathing disturbances without EEG arousal is associated with sleepiness, this observation may reflect the fact cortical activations are not 100% associated with disturbances in sleep homeostasis.

The ArI predicted by the proposed method exhibited a strong association to the MSL, with an average decrease in the MSL of 40 seconds for each doubling in the ArI ($\beta$ = -0.67, $p$ = 0.0075). This result is in concordance with similar regression models of MSL [51, 48, 49]. The regression analysis also included median arousal duration with associations of BDs and PLMs. The Me-Spon-A-Dur variable was the only significant arousal duration measure ($\beta = -1.5$, $p = 0.0028$), while Spon-ArI was insignificant. This suggests again that all arousals (spontaneous versus associated with SDB or PLMs) are not created equal with respect to their effects on daytime sleepiness. Bigger samples would be needed to examine this question more thoroughly. These results suggest that the proposed scoring system have clinical applications, as arousal variables show a significant link to MSL.

The regression analysis presented in this chapter incorporated scoring of arousals, wake, breathing disturbances, blood oxygen desaturations, and leg movements, but could be expanded by including sleep stages and lights on/off annotations. However, it is already difficult to compare significant explanatory variables due to the small effect size of each variable. This could partly be sorted by simply using much more data, although this is difficult due to the lack of available databases that contains nocturnal PSGs with associated MSLT data. Alternatively, if the sole purpose is to predict MSL, then a deep learning framework could be implemented to directly model MSL from an entire PSG. Unfortunately, however, large datasets with PSGs and MSLTs are nonexistent, and to our knowledge the WSC is the only available large sample with such data.

## 5 Conclusion

A fully automatic method for concurrent detection of arousals and wake based on convolutional and LSTM neural networks was presented. The model was trained on a dataset of 2889 PSGs and the performance was evaluated on 1026 PSGs. These PSGs came from four distinct cohorts that used different hardware for PSG recordings, ensuring robustness of the MAD detector. The model predicted arousal events with a precision of 0.72, recall of 0.81, and a F1 score of 0.76. Wake was predicted on the test set with an accuracy of 0.95. The arousal event scoring performance of the model was compared to that of 9 individual human scorers with respect to a pseudo-consensus scoring. The comparison showed that the model outperformed the average human scorer with a difference in F1 score of 0.09 on 30 PSGs from both the Wisconsin Sleep Cohort and Stanford Sleep Cohort. The robust performance on a dataset of this size suggests that the proposed method is best available method for fully automatic detection of arousals.

The arousal index predicted by MAD on an additional 1447 PSGs from the Wisconsin Sleep Cohort was compared to an associated MSLT through statistical analysis. The arousal index showed a significant association to the mean sleep latency with an average decrease in mean sleep latency of 40 seconds for each doubling in the arousal index ($\beta = -0.67$, $p = 0.0075$). An increase in the median duration of spontaneous arousal was also associated with a decrease in the mean sleep latency ($\beta = -1.5$, $p = 0.0028$). The model predictions were correlated with the mean sleep latency, showing that the model has clinical applications as a fully automated scoring tool and a pre-diagnostic tool of daytime sleepiness.

**Declaration of interest:** Dr. Mignot has received funding from Jazz pharmaceutical and has shares in Rythm, a company doing a consumer portable EEG device, but these involvements are unrelated to this project. Katie L. Stone has received grant funding from Merck, but this is unrelated to this project and they are not involved.

**Acknowledgment:** Andreas Brink-Kjaer, Alexander Neergaard Olesen and Emmanuel Mignot are partially funded by the Klarman Family Foundation. Poul Jennum is supported by internal funding from Rigshospitalet. Additional support was provided to Andreas Brink-Kjaer by the Marie and M.B. Richters, Vera and Carl Johan Michaelsens, Froeken Marie Maanssons, Oticon, Dansk Tennis, Julie Damms, and IDA foundations.

Wisconsin Sleep Cohort polysomnography data collection was supported by the US National Institutes of Health (NIH) grants 1R01AG036838, R01HL62252, and 1UL1RR02501.

The MrOS Study was funded by: U01s AG027810, AG042124, U01 AG042139,40, 43 and 45, AG042168, U01 AR066160, UL1 TR000128 and R01s HL070837-39, 40-42, 48.

Furthermore, we would like to thank The National Sleep Research Resource for offering free access to large collections of data. The NSRR is supported by Grant Number HL114473 from the National Heart, Lung, and Blood Institute, NIH.

**References**

[1] M. H. Kryger, T. Roth and W. C. Dement, Principles and Practice of Sleep Medicine (Fifth Edition), Fifth Edition ed., Philadelphia: W.B. Saunders, 2011.

[2] C. B. Saper, T. E. Scammell and J. Lu, "Hypothalamic regulation of sleep and circadian rhythms," *Nature,* vol. 437, pp. 1257-1263, 27 10 2005.

[3] J. B. Stephansen, A. N. Olesen, M. Olsen, A. Ambati, E. B. Leary, H. E. Moore, O. Carrillo, L. Lin, F. Han, H. Yan and others, "Neural network analysis of sleep stages enables efficient diagnosis of narcolepsy," *Nature communications,* vol. 9, no. 1, p. 5229, 6 December 2018.

[4] R. B. Berry, R. Brooks, C. E. Gamaldo, S. M. Harding, R. M. Lloyd, S. F. Quan, M. M. Troester, C. L. Marcus, B. V. Vaughn and S. M. Thomas, "The AASM Manual for the Scoring of Sleep and Associated Events: Rules, Terminology and Technical Specifications," *American Academy of Sleep Medicine,* 2017.

[5] P. Halász, M. Terzano, L. Parrino and R. Bódizs, "The nature of arousal in sleep," *Journal of sleep research,* vol. 13, pp. 1-23, 2004.

[6] L. J. Findley, M. E. Unverzagt and P. M. Suratt, "Automobile Accidents Involving Patients with Obstructive Sleep Apnea," *American Review of Respiratory Disease,* vol. 138, pp. 337-340, 1988.

[7] M. H. Bonnet, K. Doghramji, T. Roehrs, E. J. Stepanski, S. H. Sheldon, A. S. Walters, M. Wise and A. L. Chesson, "The scoring of arousal in sleep: reliability, validity, and alternatives," *J Clin Sleep Med,* vol. 13, 3 2007.

[8] F. De Carli, L. Nobili, P. Gelcich and F. Ferrillo, "A Method for the Automatic Detection of Arousals During Sleep," vol. 22, pp. 561-72, 9 1999.

[9] T. Sugi, F. Kawana and M. Nakamura, "Automatic EEG arousal detection for sleep apnea syndrome," *Biomedical Signal Processing and Control,* vol. 4, pp. 329-337, 2009.

[10] S. Cho, J. Lee, H. Park and K. Lee, "Detection of arousals in patients with respiratory sleep disorders using a single channel EEG," *Conf Proc IEEE Eng Med Biol Soc,* vol. 3, pp. 2733-2735, 2005.

[11] D. Popovic, M. Khoo and P. Westbrook, "Automatic scoring of sleep stages and cortical arousals using two electrodes on the forehead: Validation in healthy adults," *Journal of Sleep Research,* vol. 23, 12 2013.

[12] S. S. Shahrbabaki, C. Dissanayaka, C. R. Patti and D. Cvetkovic, "Automatic detection of sleep arousal events from polysomnographic biosignals," *2015 IEEE Biomedical Circuits and Systems Conference,* 2015.

[13] G. L. Sorensen, P. Jennum, J. Kempfner, M. Zoetmulder and H. B. D. Sorensen, "A Computerized Algorithm for Arousal Detection in Healthy Adults and Patients With Parkinson Disease," *Journal of Clinical Neurophysiology,* vol. 29, pp. 58-64, 2 2012.

[14] I. Fernández-Varela, D. Alvarez-Estevez, E. Hernández-Pereira and V. Moret-Bonillo, "A simple and robust method for the automatic scoring of EEG arousals in polysomnographic recordings," *Computers in Biology and Medicine,* vol. 87, pp. 77-86, 2017.

[15] O. Shmiel, T. Shmiel, Y. Dagan and M. Teicher, "Data mining techniques for detection of sleep arousals," *Journal of Neuroscience Methods,* vol. 179, pp. 331-337, 2009.

[16] D. C. Wallant, V. Muto, G. Gaggioni, M. Jaspar, S. L. Chellappa, C. Meyer, G. Vandewalle, P. Maquet and C. Phillips, "Automatic artifacts and arousals detection in whole-night sleep EEG recordings," *Journal of Neuroscience Methods,* vol. 258, pp. 124-133, 2016.

[17] Y. LeCun, Y. Bengio and G. Hinton, "Deep learning," *Nature,* vol. 521, pp. 436 EP -, 27 5 2015.

[18] D. A. Dean, A. L. Goldberger, R. Mueller, M. Kim, M. Rueschman, D. Mobley, S. S. Sahoo, C. P. Jayapandian, L. Cui, M. G. Morrical and others, "Scaling Up Scientific Discovery in Sleep Medicine: The National Sleep Research Resource," *Sleep,* vol. 39, pp. 1151-1164, 5 2016.

[19] J. B. Blank, P. M. Cawthon, M. L. Carrion-Petersen, L. Harper, J. P. Johnson, E. Mitson and R. R. Delay, "Overview of recruitment for the osteoporotic fractures in men study (MrOS)," *Contemp Clin Trials,* vol. 26, pp. 557-568, 10 2005.

[20] E. Orwoll, J. B. Blank, E. Barrett-Connor, J. Cauley, S. Cummings, K. Ensrud, C. Lewis, P. M. Cawthon, R. Marcus, L. M. Marshall, J. McGowan, K. Phipps, S. Sherman, M. L. Stefanick and K. Stone, "Design and baseline characteristics of the osteoporotic fractures in men (MrOS) study--a large observational study of the determinants of fracture in older men," *Contemp Clin Trials,* vol. 26, pp. 569-585, 10 2005.

[21] T. Blackwell, K. Yaffe, S. Ancoli-Israel, S. Redline, K. E. Ensrud, M. L. Stefanick, A. Laffan, K. L. Stone and Osteoporotic Fractures in Men (MrOS) Study Group, "Associations of Sleep Architecture and Sleep Disordered Breathing with Cognition in Older Community-Dwelling Men: The MrOS Sleep Study," *J Am Geriatr Soc,* vol. 59, pp. 2217-2225, 07 12 2011.

[22] S. Redline, P. V. Tishler, T. D. Tosteson, J. Williamson, K. Kump, I. Browner, V. Ferrette and P. Krejci, "The familial aggregation of obstructive sleep apnea.," *American Journal of Respiratory and Critical Care Medicine,* vol. 151, pp. 682-687, 1995.

[23] S. Redline, P. V. Tishler, M. Schluchter, J. Aylor, K. Clark and G. Graham, "Risk Factors for Sleep-disordered Breathing in Children," *American Journal of Respiratory and Critical Care Medicine,* vol. 159, pp. 1527-1532, 1999.

[24] P. Peppard, *The Wisconsin Sleep Cohort.*

[25] T. Young, L. Finn, P. E. Peppard, M. Szklo-Coxe, D. Austin, F. J. Nieto, R. Stubbs and K. M. Hla, "Sleep Disordered Breathing and Mortality: Eighteen-Year Follow-up of the Wisconsin Sleep Cohort," *Sleep,* vol. 31, pp. 1071-1078, 01 8 2008.

[26] O. Andlauer, H. Moore, L. Jouhier, C. Drake, P. E. Peppard, F. Han, S.-C. Hong, F. Poli, G. Plazzi, R. O'Hara and others, "Nocturnal rapid eye movement sleep latency for identifying patients with narcolepsy/hypocretin deficiency," *JAMA neurology,* vol. 70, pp. 891-902, 2013.

[27] M. Bonnet, D. Carley, M. Carskadon, P. Easton, C. Guilleminault, R. Harper, B. Hayes, M. Hirshkowitz, P. Ktonas, S. Keenan and others, "ASDA report: EEG arousals: scoring rules and examples," *Sleep,* vol. 15, pp. 173-184, 1992.

[28] J. A. Hobson, "A manual of standardized terminology, techniques and scoring system for sleep stages of human subjects: A. Rechtschaffen and A. Kales (Editors).(Public Health Service, US Government Printing Office, Washington, DC, 1968, 58 p., $ 4.00)," *Electroencephalography and clinical neurophysiology,* vol. 26, p. 644, 1969.

[29] H. Moore, E. Leary, S.-Y. Lee, O. Carrillo, R. Stubbs, P. Peppard, T. Young, B. Widrow and E. Mignot, "Design and Validation of a Periodic Leg Movement Detector," *PLOS ONE,* vol. 9, pp. 1-30, 12 2014.

[30] P. He, G. Wilson and C. Russell, "Removal of ocular artifacts from electro-encephalogram by adaptive filtering," *Medical and Biological Engineering and Computing,* vol. 42, pp. 407-412, 01 5 2004.

[31] Y. LeCun, L. Bottou, G. B. Orr and K. R. Müller, "Efficient BackProp," in *Neural Networks: Tricks of the Trade*, G. B. Orr and K. Müller, Eds., Berlin, Heidelberg: Springer Berlin Heidelberg, 1998, pp. 9-50.

[32] Z. C. Lipton, D. C. Kale, C. Elkan and R. C. Wetzel, "Learning to Diagnose with LSTM Recurrent Neural Networks," *International Conference on Learning Representations,* 2016.

[33] S. Biswal, H. Sun, B. Goparaju, M. B. Westover, J. Sun and M. T. Bianchi, "Expert-level sleep scoring with deep neural networks," *Journal of the American Medical Informatics Association,* vol. 25, no. 12, pp. 1643-1650, 2018.

[34] A. Supratak, H. Dong, C. Wu and Y. Guo, "DeepSleepNet: A model for automatic sleep stage scoring based on raw single-channel EEG," *IEEE Transactions on Neural Systems and Rehabilitation Engineering,* vol. 25, pp. 1998-2008, 2017.

[35] S. Hochreiter and J. Schmidhuber, "Long Short-Term Memory," *Neural Comput.,* vol. 9, pp. 1735-1780, 11 1997.

[36] A. N. Olesen, P. Jennum, P. Peppard, E. Mignot and H. B. Sorensen, "Deep residual networks for automatic sleep stage classification of raw polysomnographic waveforms," *2018 40th Annual International Conference of the IEEE Engineering in Medicine and Biology Society (EMBC),* pp. 1-4, 29 October 2018.

[37] K. He, X. Zhang, S. Ren and J. Sun, "Deep Residual Learning for Image Recognition," *Proceedings of the IEEE International Conference on Computer Vision (IICV),* pp. 1026-1034, 2015.

[38] S. Ioffe and C. Szegedy, "Batch Normalization: Accelerating Deep Network Training by Reducing Internal Covariate Shift," *Proceedings of the 32nd International Conference on International Conference on Machine Learning,* vol. 37, pp. 448-456, 2015.

[39] G. E. Dahl, T. N. Sainath and G. E. Hinton, "Improving deep neural networks for LVCSR using rectified linear units and dropout," in *Acoustics, Speech and Signal Processing (ICASSP), 2013 IEEE International Conference on*, 2013.

[40] D. P. Kingma and J. Ba, "Adam: A Method for Stochastic Optimization," *Proc. 3rd Int. Conf. Learn. Representations,* 2014.

[41] X. Glorot and Y. Bengio, "Understanding the difficulty of training deep feedforward neural networks," in *In Proceedings of the International Conference on Artificial Intelligence and Statistics (AISTATS'10). Society for Artificial Intelligence and Statistics*, 2010.

[42] H. Koch, L. Douglas Schneider, L. A Finn, E. B Leary, P. E Peppard, E. Hagen, H. Bjarup Dissing Sorensen, P. Jennum and E. Mignot, "Breathing Disturbances Without Hypoxia Are Associated With Objective Sleepiness in Sleep Apnea," *SLEEP,* vol. 40, 9 2017.

[43] M. Manconi, I. Zavalko, C. L. Bassetti, E. Colamartino, M. Pons and R. Ferri, "Respiratory-Related Leg Movements and Their Relationship with Periodic Leg Movements During Sleep," *Sleep,* vol. 37, no. 3, pp. 497-504, 1 March 2014.

[44] S. Aritake, T. Blackwell, K. W. Peters, M. Rueschman, D. Mobley, M. G. Morrical, S. F. Platt, T.-T. L. Dam, S. Redline, Osteoporotic Fractures in Men (MrOS) Study Research and others, "Prevalence and associations of respiratory-related leg movements: the MrOS sleep stud," *Sleep medicine,* vol. 16, no. 10, pp. 1236-1244, 30 June 2015.

[45] S. L. Zeger and K.-Y. Liang, "Longitudinal data analysis for discrete and continuous outcomes," *Biometrics,* pp. 121-130, 1986.

[46] S. Ratcliffe and J. Shults, "GEEQBOX: A MATLAB Toolbox for Generalized Estimating Equations and Quasi-Least Squares," *Journal of Statistical Software, Articles,* vol. 25, pp. 1-14, 2008.

[47] H. Sun, J. Jia, B. Goparaju, G.-B. Huang, O. Sourina, M. T. Bianchi and M. B. Westover, "Large-scale automated sleep staging," *Sleep,* vol. 40, 2017.

[48] S. E. Martin, H. M. Engleman, R. N. Kingshott and N. J. Douglas, "Microarousals in patients with sleep apnoea/hypopnoea syndrome," *Journal of Sleep Research,* vol. 6, pp. 276-280, 1997.

[49] T. Roehrs, F. Zorick, R. Wittig, W. Conway and T. Roth, "Predictors of Objective Level of Daytime Sleepiness in Patients with Sleep-Related Breathing Disorders," *CHEST,* vol. 95, pp. 1202-1206, 1989.

[50] R. D. Chervin, "Periodic leg movements and sleepiness in patients evaluated for sleep-disordered breathing," *American journal of respiratory and critical care medicine,* vol. 164, pp. 1454-1458, 2001.

[51] P. H. Leng, S. Y. Low, A. Hsu and S. F. Chong, "The clinical predictors of sleepiness correlated with the multiple sleep latency test in an Asian Singapore population," *Sleep,* vol. 26, pp. 878-881, 2003.